\documentclass[entropy,article,accept,pdftex,moreauthors]{Definitions/mdpi} 
\firstpage{1} 
\pubvolume{1}
\issuenum{1}
\articlenumber{0}
\pubyear{2025}
\copyrightyear{2025}
\externaleditor{Roberto Zivieri} 
\datereceived{7 August 2025} 
\daterevised{15 September 2025} 
\dateaccepted{18 September 2025} 
\datepublished{ } 
\hreflink{https://doi.org/} 

\usepackage{bm}        

\usepackage{physics}

\usepackage{mathtools}

\usepackage{soul}

\newcommand{\half}{{\textstyle\frac{1}{2}}}
\newcommand{\quarter}{{\textstyle\frac{1}{4}}}

\newcommand{\twelth}{{\textstyle\frac{1}{12}}}

\newcommand{\imi}{\mathrm{i}}

\newcommand{\Boltz}{ k_{\rm\scriptscriptstyle B}}
\newcommand{\Hil}{{\mathcal H}}
\newcommand{\J}{{J\vphantom{\overline{J}}}}
\newcommand{\Jbar}{{\overline{J}}}
\newcommand{\acommInline}[2]{\{#1,#2\}}
\newcommand{\D}{\text{d}}
\newcommand{\Bln}{{\text{Bln}}}
\newcommand{\Bell}{{\text{\tiny Bell}}}

\definecolor{rkrPurple}{HTML}{73024F}

\Title{\textls[-10]{No-Signaling in Steepest Entropy Ascent: A Nonlinear, Non-Local, Non-Equilibrium Quantum Dynamics of Composite Systems Strongly Compatible with the Second Law}}

\TitleCitation{No-Signaling in Steepest Entropy Ascent: A Nonlinear, Non-Local, Non-Equilibrium Quantum Dynamics of Composite Systems Strongly Compatible with the Second Law}

\Author{Rohit Kishan Ray 
 $^{1,2,}$*\orcidA{} and Gian Paolo Beretta $^{3}$\orcidB{}}

\AuthorNames{Rohit Kishan Ray and Gian Paolo Beretta}

        \AuthorCitation{Ray, R.K.; Beretta, G.P.}

\address{%
$^{1}$ \quad Department of Material Science and Engineering, Virginia Tech, Blacksburg, VA 24061, USA\\
$^{2}$ \quad Center for Theoretical Physics of Complex Systems, Institute for Basic Science (IBS), \mbox{Daejeon 34126, Republic of Korea}\\
$^{3}$ \quad Department of Mechanical and Industrial Engineering, University of Brescia, 25123 Brescia, Italy; gianpaolo.beretta@unibs.it}

\corres{Correspondence: rkray@vt.edu}

\abstract{Lindbladian formalism models open quantum systems using a `bottom-up' approach, deriving linear dynamics from system--environment interactions. We present a `top-down' approach starting with phenomenological constraints, focusing on  a system's structure, subsystems' interactions, and environmental effects and often using a non-equilibrium variational principle designed to enforce strict thermodynamic consistency. However, incorporating the second law’s requirement---that Gibbs states are the sole stable equilibria---necessitates nonlinear dynamics, challenging no-signaling principles in composite systems. We reintroduce `local perception operators' and show that they allow to model signaling-free non-local effects. Using the steepest-entropy-ascent variational principle as an example, we demonstrate the validity of the `top-down' approach for integrating quantum mechanics and thermodynamics in phenomenological models, with potential applications in quantum computing and resource theories.}


\begin{document}
\section{\label{sec:intro}Introduction}
The study of open quantum systems broadly falls into two distinct approaches. The dominant method involves the Gorini--Kossakowski--Lindblad--Sudarshan (GKLS) master equation, typically under the assumption of weak coupling between a quantum system and its environment (for recent reviews, see~\cite{kosloff_2019_quantum, chruscinski_2022_dynamicalb}).
This approach derives thermodynamically consistent reduced system dynamics starting from the Schr\"{o}dinger equation combined system-environment evolution and subsequently tracing out environmental degrees of freedom~\cite{schmidt_2023_stochastic,riera-campeny_2021_quantum,manzano_2021_thermodynamics}. We refer to this widely adopted strategy as the `bottom-up' approach. Its effectiveness relies on idealized assumptions regarding the environment and the careful construction of appropriate Kraus operators, and it typically presumes that the initial state of the system-environment composite is factorizable.

In contrast, an alternative yet less explored methodology exists: the phenomenological or `top-down' approach. This paradigm directly posits a general local dynamical law for the system's density operator based primarily on thermodynamic and phenomenological considerations. It inherently accommodates strongly coupled subsystems and composite structures. Crucially, it explicitly incorporates the thermodynamic requirement that Gibbs states are the only conditionally stable equilibrium states~\cite{simmons_1981_essentiala, beretta_1986_theorema}, necessitating a departure from strictly linear evolution. Simmons and Park~\cite{simmons_1981_essentiala} demonstrated early on that nonlinearity is essential to achieve a consistent integration of quantum mechanics and thermodynamics. It is important to clarify here that our approach is distinct from the nonlinear quantum mechanics (NQM) recently subjected to stringent experimental tests~\cite{polkovnikov_2023_experimental,melnychuk_2024_improved}; rather, it represents a phenomenological route toward thermodynamically consistent, irreversible and non-equilibrium quantum dynamical models.

Recent experimental advances underscore the practical necessity for top-down phenomenological models, especially in scenarios where detailed environmental modeling is inaccessible or impractical~\cite{somhorst_2023_quantum}. Such models naturally describe `apparent decoherence' stemming from unknown environmental interactions, employing minimal empirical parameters. Moreover, they conveniently handle situations where multiple conserved quantities, possibly non-commuting, are relevant at the local level~\cite{YungerHalpern_naturecomm_2016}.

Among the existing top-down formalisms, the Steepest Entropy Ascent (SEA) framework has emerged as particularly powerful and conceptually robust. Originating from the foundational works of Hatsopoulos and Gyftopoulos~\cite{hatsopoulos_1976_unified,hatsopoulos_1976_unifieda,hatsopoulos_1976_unifiedb,hatsopoulos_1976_unifiedc}, and subsequently developed extensively by Beretta~\cite{beretta_1984_quantuma,beretta_1985_quantuma, maddox_1985_uniting,pelegrin_1987_colloque,gheorghiu-svirschevski_2001_nonlinear, gheorghiu-svirschevski_2001_addendum, beretta_2005_nonlinear, beretta_2006_nonlinearb,beretta_2009_nonlinear, beretta_2010_maximum}, SEA integrates thermodynamic irreversibility into quantum dynamics from first principles. This formalism has been successfully employed across diverse quantum systems~\cite{cano-andrade_2015_steepestentropyascent,vonspakovsky_2014_trends,ray_2022_steepestb}, including quantum computational scenarios~\cite{tabakin_2017_modela,tabakin_2023_locala}. Nevertheless, despite its robustness and conceptual appeal, a rigorous demonstration of SEA's compatibility with fundamental quantum principles, particularly the no-signaling condition, has thus far remained underexplored~\cite{beretta_2005_nonlinear,beretta_2009_nonlinear,beretta_2010_maximum,vonspakovsky_2014_trends}.

This paper explicitly fills this critical gap. We rigorously prove that SEA dynamics inherently satisfy the no-signaling condition, even though its equations of motion incorporate nonlinearities and nonlocal structures. To this end, we introduce and utilize Local Perception Operators (LPOs), a conceptual tool that ensures subsystem locality in a nonlinear dynamical context. We establish the invariance of LPOs under local unitary transformations, which guarantees that no signaling between non-interacting subsystems emerges from the SEA dynamics. We further substantiate our claims through nontrivial illustrative examples. While the primary objective is to demonstrate that SEA-type dynamics respect no-signaling, we also take this opportunity to clarify the foundational motivation for invoking the SEA framework in the first place. In doing so, we show how it resolves subtle but important paradoxes that often go unaddressed in conventional \mbox{bottom-up approaches.}

The paper is structured as follows. Section~\ref{sec:nosignaling} formally defines signaling and explores its relationship with nonlinearity. In Section~\ref{sec:philosophy}, we revisit the philosophical motivations that led to the inception of SEA dynamics. Section~\ref{sec:ontic_state} discusses conceptual issues surrounding quantum state individuality that nonlinear dynamics help resolve. Section~\ref{sec:measure_representation} introduces a measure-theoretic representation for mixed ensembles, facilitating the rigorous handling of nonlinear dynamics. The Local Perception Operators (LPOs) are defined and analyzed in Section~\ref{sec:lpo}, and their critical role in establishing no-signaling is rigorously demonstrated in Section~\ref{sec:nosignaling_lpo}. The composite-system SEA equation is developed structurally in Section~\ref{sec:composite_SEA}, followed by a variational derivation in Sections~\ref{sec:variational} and~\ref{sec:SEAsimple}. Section~\ref{sec:sea_example} presents explicit numerical examples, and Section~\ref{sec:conclusion} summarizes our findings.

\section{\label{sec:nosignaling}No-Signaling and Nonlinearity}

Quantum mechanics (QM), as described by the Schr\"{o}dinger-von Neumann formalism, is traditionally characterized by its linearity in state space and time evolution. Mean values of global and local properties and their time evolutions under unitary (and completely positive) dynamics  are linear functionals of the density operator. This linearity implies no-cloning, first shown by \citet{park_1970_concept} and later rediscovered by \citet{ghirardi_2013_entanglement}, \citet{wootters_1982_single}, and \citet{dieks_1982_communication}. Linearity also implies no-signaling~\cite{eberhard_1978_bells,simon_2001_nosignaling,bona_2003_comment,simon_2003_simon}. However, Weinberg questioned this linear characteristic of QM~\cite{weinberg_1989_testing}. He introduced nonlinearity into the operator formalism of the Schr\"{o}dinger equation while preserving order-one homogeneity. This was one of the earliest examples of what is now broadly referred to as nonlinear quantum mechanics (NQM). \citet{gisin_1990_weinberg} showed that such formalism could facilitate faster-than-light communication, \\emph{viz.}, 
 signaling. Polchinski~\cite{polchinski_1991_weinberg} proposed an alternative method to prevent signaling by restricting local evolutions in order to depend solely on the corresponding reduced density operators. However, this formalism still allowed signaling between different branches of the wave function (Everett telephone)~\cite{polchinski_1991_weinberg}. \mbox{\citet{wodkiewicz_1990_weinberg}} analyzed the nonlinear evolution of two-level atoms, with solutions later shown to depend on interpretation~\cite{czachor_1996_nonlinear}. Czachor also showed that a nonlinear operator induces mobility phenomena (non-conservation of the inner-product of two pure states) \cite{czachor_1991_mobility}. 

Nonlinearity introduced via stochastic QM through Lindblad operator formalism~\cite{gisin_1995_relevant} for open quantum systems has been shown to respect no-signaling. Weinberg-inspired nonlinearities, interestingly, enable faster algorithms to solve \emph{NP}-complete problems in polynomial time~\cite{abrams_1998_nonlinear}. Later work~\cite{ferrero_2004_nonlinear} demonstrated that nonlinearity in QM can be incorporated without violating no-signaling, as long as time evolution is nonlinear while state space and operators remain linear. More recently, convex quasilinear maps~\cite{rembielinski_2020_nonlinear,rembielinski_2021_nonlinear} were shown to support nonlinear QM dynamics without signaling, preserving key features of QM. \citet{rembielinski_2020_nonlinear} identified this as the minimal allowable deviation from QM's linear structure. In a recent work, \citet{kaplan_2022_causal} showed that, in a low-temperature limit, the non-linearity introduced in quantum field theory results in causal nonlinear quantum mechanics, which is also another version of NQM.

\begin{figure}[H] 
    \includegraphics[height=0.65\columnwidth]{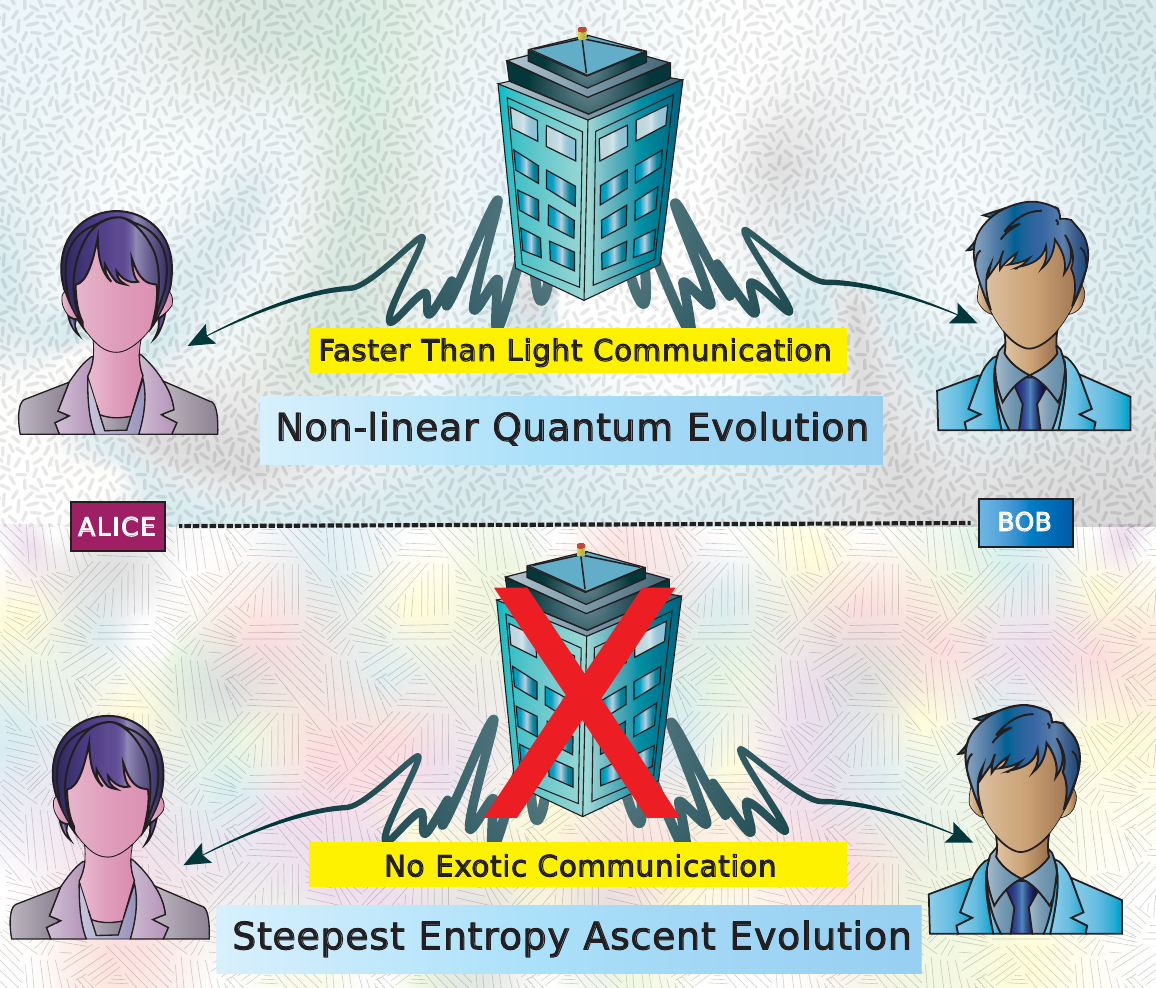}
    \caption{Schematic representation of no-signaling situation in the case of nonlinear quantum evolution. Alice and Bob are correlated but are non-interacting in a nonlinear quantum theory (we imply the nonlinearities via the texture of the embedding background in the schematic). In the top part,  a non-linear quantum evolution as described by Polchinski~\cite{polchinski_1991_weinberg} or Gisin~\cite{gisin_1990_weinberg} may entail faster-than-light communication in some form (denoted as the telephone booth; see text for details). In the bottom part, instead, the embedding nonlinear steepest-entropy-ascent evolution cannot establish a similar exotic communication.}
    \label{fig:schematics}
\end{figure}

Therefore, introducing nonlinearity in quantum mechanics (QM) can lead to exotic interactions that are difficult to justify physically. Nonlinear theories, as inspired by the Weinberg approach, allow for entropy oscillation~\cite{czachor_1991_mobility}. The stochastic, jump-induced mixing of pure states~\cite{gisin_1995_relevant} offers a basis for developing a thermodynamically consistent nonlinear master equation for the density operator of an open system. However, the search for theoretically consistent nonlinear models of quantum thermodynamics remains an open challenge. This is where we claim---and later demonstrate in the following \mbox{sections---that} the `top-down' approach of SEA resolves the signaling problem entirely (pictorially sketched in Figure \ref{fig:schematics}). From a philosophical perspective, SEA also addresses other conceptual conundrums, in particular the Schr\"{o}dinger-Park paradox~\cite{schrodinger_1936_probability,park_1968_nature}, as was already discussed by \mbox{\citet{park_1988_thermodynamic}} and \citet{beretta_2006_hatsopoulos}.

The no-signaling condition, as noted in~\cite{ferrero_2004_nonlinear}, is typically enforced by requiring that, in the absence of mutual interactions between subsystems A and B, the evolution of A's local observables depends solely on its reduced state. Formally, we express this as follows:
\begin{equation}\label{eq:fererro_nosignaling}
    \dv{\rho_J}{t} = f(\rho_J),
\end{equation}
where $\rho_J$ is the reduced density operator (local state) of subsystem $J$. The SEA formalism, however, adopts a less restrictive perspective~\cite{beretta_2005_nonlinear}. It requires that, if A and B are non-interacting, the law of evolution must not permit a local operation within B to influence the time evolution of A's local (reduced, marginal) state. Consequently, the condition   $\rho_A=\rho_A'$, applied to the two different states $\rho\ne\rho_A{\otimes}\rho_B$ and $\rho'=\rho_A{\otimes}\rho_B$, does not necessitate {${\rm{d}}\rho_A/{\rm{d}}t={\rm{d}}\rho_A'/{\rm{d}}t$}. This is because the local evolution can still be influenced by past interactions, such as existing entanglement and correlations, without violating the no-signaling principle. This highlights two key ideas: (1) analyzing local evolutions allows detection of correlations, but only those that can be classically communicated between subsystems, and (2) in the absence of interactions, nonlinear dynamics can cause correlations to diminish (spontaneous decoherence) but cannot create new correlations. We formally express this no-signaling condition as follows (see Section \ref{sec:composite_SEA} for details),
\begin{equation}\label{eq:no_signaling_condition}
	\dv{\rho_J}{t} = f(\rho_J,(C_k)^J),
\end{equation}
where $(C_k)^J$ are `local perception operators' (LPO) (see Section \ref{sec:lpo} for their definition).

\section{\label{sec:philosophy}Philosophical Motivations Leading to SEA}

Philosophically, SEA evolution was designed to conceive the irreversible equilibration and decoherence as a fundamental spontaneous dynamical feature, contrary to the coarse-graining approach. Entropy, the second law, and irreversibility could attain a more fundamental status as deterministic outcomes of quantal evolution~\cite{beretta_1986_intrinsic}, without contradicting standard QM. From the outset~\cite{beretta_1984_quantuma,beretta_1985_entropy,beretta_1985_quantuma}, nonlinearity was recognized as essential~\cite{simmons_1981_essentiala,park_1983_knotsa} for this purpose. While the prevalent notion was---and still is---that the second law is emerging and  statistical, the pioneers of SEA believed that the many `knots of thermodynamics'~\cite{park_1983_knotsa,hatsopoulos_2008_where} could, at least conceptually, be untied by elevating it to a more fundamental stature. This motivated Hatsopoulos and Gyftopoulos (HG) to develop a unified theory of mechanics and thermodynamics~\cite{hatsopoulos_1976_unified,hatsopoulos_1976_unifieda,hatsopoulos_1976_unifiedb,hatsopoulos_1976_unifiedc}. They laid the foundation for what is now known as resource theory in quantum thermodynamics by extending ontological quantum states to include both pure and mixed density operators. The HG unified (resource) theory explicitly defines and fully characterizes the concept---later rediscovered as `ergotropy'~\cite{Allahverdyan_2004}---of maximum work extractable in a unitary process with cyclic parameter changes (theorems 3.2 and 3.3 in~\cite{hatsopoulos_1976_unifieda}, which pioneered also the use of `majorization' in quantum theory). 

Beretta's doctoral thesis~\cite{beretta_1981_thesis} aimed to complete the HG theory with two key goals: (1) to design an equation of motion that establishes the second law of thermodynamics as a theorem, ensuring Gibbs states are the only conditionally stable equilibrium states; and (2) to develop an unambiguous statistical representation of heterogeneous ensembles formed by the statistical mixing of homogeneous (pure) ensembles as one-to-one with the full set of pure and mixed density operators. 
  
The outcome of the first objective was the SEA formalism, including the introduction of local perception operators. SEA formalism proved to be both robust and versatile, making it a powerful tool for thermodynamically consistent non-equilibrium modeling. Its applications expanded beyond its original scope to include phenomenological models in areas such as decoherence~\cite{montanez-barrera_2022_Decoherencepredictions}, mesoscopic transport rheology~\cite{vonspakovsky_2016,vonspakovsky_2018}, phase transitions~\cite{vonspakovsky_2019}, and quantum computing~\cite{tabakin_2017_modela,ray_2022_steepestb,vonspakovsky_2020,tabakin_2023_locala}.

 The second objective led to the representation of generic ensembles using measure-theoretic distributions  over the set of ontological states.  In the context of
 quantum thermodynamics, an ``unambiguous statistical representation of heterogeneous ensembles'' ensures a clear distinction between intrinsic quantum uncertainties and classical statistical mixing. This approach---detailed in Section~\ref{sec:measure_representation} following \cite{beretta_1981_thesis,beretta_2006_hatsopoulos}---represents heterogeneous ensembles as distributions over the full set of pure and mixed density operators, thus avoiding ambiguities associated with von Neumann's statistical interpretation and resolving the Schr\"{o}dinger-Park paradox.

\section{\label{sec:ontic_state}Nonlinear Dynamics and the Ontic State Conundrum}

From a foundational perspective, the HG approach of assigning ontological status to the full set of density operators, pure and mixed, offers a potential resolution to the Schr\"{o}dinger--Park paradox concerning 
the ambiguity of individual quantum states~\cite{schrodinger_1936_probability,park_1968_nature,beretta_2006_hatsopoulos}. As noted by \citet{park_1988_thermodynamic}, treating density operators as one-to-one with homogeneous ensembles resolves ambiguity under two conditions: (1) that they evolve unitarily for short times, with nonlinear dynamics dominating over longer periods; and  (2) that a full tomography, even for heterogeneous preparations, must include both linear and nonlinear observables, such as entropy, adiabatic availability, ergotropy, and free energy.

The von Neumann prescription \cite{vonneumann_2018_mathematical} {(Chapter III)} 
 assigns a density operator to a statistical mixture of pure states, with mixing coefficients reflecting ignorance. However, \citet{schrodinger_1936_probability} questioned this from the outset due to the non-unique decomposition of a mixed density operator into weighted sums of pure states. von Neumann's approach irrecoverably blends quantal probabilities with non-quantal statistical weights, creating the ambiguity highlighted by Schr\"{o}dinger and Park. To understand how a law of nonlinear evolution may resolve this issue, let $\hat\Pi^0_1$ and $\hat\Pi^0_2$ be two distinct homogeneous preparations. At time \(t=0\), they generate two initial ensembles of identically prepared, strictly isolated, and non-interacting systems. The `hat' on $\Pi$ denotes the homogeneity of the preparation, and superscript `0' refers to time $t=0$. According to the traditional statistical formulation by von Neumann, a preparation, $\hat\Pi$, is {\textit{homogeneous}} 
 if and only if it cannot be replicated using any statistical composition of two {\textit{different}} preparations. The statistics of measurement outcomes from a homogeneous preparation are represented as a pure (idempotent) density operator, $\hat\rho$, satisfying $\hat\rho = \hat\rho^2$. Consider a third (heterogeneous) preparation, $\Pi^0_3$,  obtained via the statistical mixture of $\hat\Pi^0_1$ and $\hat\Pi^0_2$ with a fixed weight, $0 < w < 1$, so that, formally, we may write $\Pi^0_3=w\,\hat\Pi^0_1+(1{-}w)\,\hat\Pi^0_2$. To obtain quantum statistical mechanics (QSM) consistent with Gibbs--Boltzmann, Fermi--Dirac, and Bose--Einstein  equilibrium distributions, von Neumann postulated that a heterogeneous preparation, such as $\Pi^0_3$, should be represented as a mixed density operator, $\rho^0_3 = w\,\hat\rho^0_1+(1{-}w)\,\hat\rho^0_2$. This formulation, partly due to the linearity of quantum mechanics, has served as the foundation for successful frameworks such as equilibrium QSM and the Lindbladian theory of open systems.\\

These successes often lead us to overlook the Schr\"{o}dinger--Park paradox or accept it as a seemingly enigmatic aspect of quantum theory. It arises when interpreting a heterogeneous ensemble. If we assume that each member of the ensemble is prepared in either state $\hat\rho^0_1$ (with probability $w$) or $\hat\rho^0_2$ (with probability $1{-}w$), excluding all other states, we face a contradiction. The same mixed density operator, $\rho^0_3$, allows infinitely many alternative decompositions. For instance, the decomposition $\rho^0_3 = w'\,\hat\rho^0_4 + (1{-}w')\,\hat\rho^0_5$ suggests that the ensemble consists only of $\hat\rho^0_4$ and $\hat\rho^0_5$, excluding $\hat\rho^0_1$ and $\hat\rho^0_2$. This inconsistency challenges the notion of individual states in a heterogeneous ensemble and complicates the interpretation of QSM.
Under a linear evolution map for the homogeneous preparations, $\hat\Pi^t=\mathcal{L}_t(\hat\Pi^0)$, and mixing of homogeneous preparations with  time-invariant  weights, we have 
\begin{equation}
\Pi_3^t=w\,\hat\Pi^t_1+(1{-}w)\,\hat\Pi^t_2=w\,\mathcal{L}_t(\hat\Pi_1^0)+(1{-}w)\,\mathcal{L}_t(\Pi_2^0)
\end{equation}
and, via the linearity of $\mathcal{L}_t$,  $\Pi_3^t= \mathcal{L}_t(\hat\Pi_3^0)$. Thus,  the von Neumann recipe gives $\rho^t_3 = w\,\hat\rho^t_1+(1{-}w)\,\hat\rho^t_2$ and, for  linear observables, it yields the correct  statistics of measurements,
\begin{equation}
\Tr (A\, \rho^t_3) = w\,\Tr (A\, \hat\rho^t_1)+(1{-}w)\,\Tr (A\, \hat\rho^t_2).
\end{equation}

It also entails that, for an $N$-level system, a full tomography of a heterogeneous preparation requires the measurement of a quorum of only $N^2{-}1$ independent linear~observables.

Instead, if the evolution map for homogeneous preparations is nonlinear, $\hat\Pi^t=\mathcal{N}_t(\hat\Pi^0)$, and the statistical mixing 
weights remain time-invariant, we obtain 
\begin{equation}
\Pi_3^t=w\,\hat\Pi^t_1+(1{-}w)\,\hat\Pi^t_2=w\,\mathcal{N}_t(\hat\Pi_1^0)+(1{-}w)\,\mathcal{N}_t(\Pi_2^0)
\end{equation}
and, in general, due to the nonlinearity of $\mathcal{N}_t$, this does not satisfy $\Pi_3^t\ne \mathcal{N}_t(\hat\Pi_3^0)$. This contradicts the von Neumann prescription, which would assign the density operator $\rho^t_3 = w\,\rho^t_1+(1{-}w)\,\rho^t_2$ to preparation $\Pi_3^t$ and $\rho^0_3 = w\,\rho^0_1+(1{-}w)\,\rho^0_2$ to preparation $\Pi_3^0$. 
As noted by \citet{rembielinski_2020_nonlinear}, a sufficient condition for a nonlinear map to be no-signaling is convex-quasilinearity. That is, it must always admit a $w'$, with $0 \leq w' \leq 1$, such that  $
\rho^t_3 = w'\,\rho^t_1 + (1{-}w')\,\rho^t_2.$ \citet{ferrero_2004_nonlinear} made a similar observation but left its interpretation as an open problem.

The HG ontological hypothesis alleviates this interpretation issue. Regardless of how a density operator is decomposed, $w$ and $1{-}w$  
no longer represent epistemic ignorance. This is because both pure and mixed density operators have ontic status, meaning they represent  
homogeneous preparations.

\section{\label{sec:measure_representation}Measure-Theoretic Representation of Statistics from Mixed Ensembles}

To resolve the Schr\"{o}dinger-Park paradox about individual states in QSM without contradicting the successes of QM, Beretta~\cite{beretta_1981_thesis} proposed replacing the von Neumann prescription for measurement statistics from mixed ensembles with an unambiguous representation akin to classical statistical mechanics. 
The recipe assigns to every preparation $\Pi$  (whether homogeneous or heterogeneous) a normalized {\textit{measure}} $ \mu_\Pi $ defined on the (ontic) quantal state domain  $ \mathcal{P} $, which consists of the mathematical objects representing homogeneous preparations (homogeneous ensembles). The normalization condition is 
\begin{equation}\label{eq:measure_normalization}
	 \mu_\Pi ( \mathcal{P} ) = \int_{\mathcal{P}} \mu_\Pi ( d \rho ) = 1.
\end{equation} 
In usual von Neumann QSM,  the quantal state domain $ \mathcal{P}_\text{QSM} $ is the set of all one-dimensional projection operators (i.e., the idempotent density operators) defined on the Hilbert space of the system. Instead, the HG-unified theory assumes the broader  quantal state domain  $\mathcal{P}_\text{HG} $ consisting of the set of all possible density operators $\rho$, idempotent \mbox{and not.}

Among the measures defined on  $ \mathcal{P} $, the \textit{{Dirac measures}}, defined as follows, play a crucial role. Let $ \mathcal{E} $ denote any subset of $ \mathcal{P} $, and then 
\begin{equation}\label{eq:Dirac_measure}
	\delta_\rho( \mathcal{E} ) =\begin{cases} 1,~~\text{ if } \rho \in \mathcal{E}\\ 0,~~\text{ if } \rho \notin \mathcal{E} \end{cases}
\end{equation}
The  support of a Dirac measure, $\delta_{\tilde\rho}$, i.e., the subset of $ \mathcal{P} $ for which it is nonzero, is a single point coinciding with the state operator $\tilde\rho$. 

The statistical mixing of preparations, $\Pi_3=w\,\Pi_1+(1{-}w)\,\Pi_2$, is represented by the weighted sum of the corresponding measures, 
\begin{equation}\label{eq:mixed_measure}
	 \mu_{\Pi_3}=w\,\mu_{\Pi_1}+(1{-}w)\,\mu_{\Pi_2}.
\end{equation}
This representation removes ambiguities because of the following:  
\begin{enumerate}
    \item No Dirac measure can be decomposed into a weighted sum of different Dirac measures, satisfying von Neumann's definition of homogeneous preparations.
    \item Any measure has a unique decomposition into a weighted sum (or integral) of Dirac measures---removing the Schr\"{o}dinger--Park paradox. This ensures that each  member of a heterogeneous ensemble is in some well defined (albeit unknown) \mbox{individual state.}
\end{enumerate}

The expected mean value of any physical observable represented by the (linear or nonlinear) functional $ g $ defined on $ \mathcal{P} $ is given for preparation $\Pi_3=w\,\Pi_1+(1{-}w)\,\Pi_2$ by
\begin{align}\label{eq:expected_value_1}
	&\overline{\langle  g \rangle}_{\Pi_3} =\int_{\mathcal{P}}  g ( \rho
	)\, \mu_{\Pi_3}( \D \rho )\nonumber\\ &= \int_{\mathcal{P}}  g ( \rho
	)\, \left[w\,\mu_{\Pi_1}( \D \rho ) +(1{-}w)\, \mu_{\Pi_2}( \D \rho )\right] \nonumber\\ &=  w\!\int_{\mathcal{P}}  g ( \rho
	)\, \mu_{\Pi_1}( \D \rho ) +(1{-}w)\!\int_{\mathcal{P}}  g ( \rho
	)\, \mu_{\Pi_2}( \D \rho ) .
\end{align}
If the component preparations are homogeneous, i.e., if $\mu_{\hat\Pi_1}=\delta_{\rho_1}$ and $\mu_{\hat\Pi_2}=\delta_{\rho_2}$, then
\begin{align}\label{eq:expected_value_2}
	&\overline{\langle  {g} \rangle}_{\Pi_3} =  w\!\int_{\mathcal{P}}  g ( \rho
	)\, \delta_{\rho_1} ( \D \rho ) +(1{-}w)\!\int_{\mathcal{P}}  g ( \rho
	)\, \delta_{\rho_2} ( \D \rho )\nonumber\\ &=w\, g(\rho_1) + (1{-}w)\,g(\rho_2)
	\nonumber\\ &=w\, \langle  {g} \rangle_{\hat\Pi_1} + (1{-}w)\,\langle  {g} \rangle_{\hat\Pi_2}\,.
\end{align}
For example, for $\mathcal{P}{ \,=\, } \mathcal{P}_\text{HG} $ and $g(\rho)=-\Boltz\Tr(\rho\ln\rho)$, this gives the `proper' expected value of measurements of the von Neumann entropy, i.e., the weighted sum of the entropies of the component homogeneous sub-ensembles.

 It is noteworthy that, for observables represented by linear functionals on $\mathcal{P}$, such as $g(\rho)=\Tr(G\rho)$,  we have
 \begin{align}\label{eq:linear_observable}
 	\overline{\langle  {g} \rangle}_{\Pi_3}&=w\, \Tr(G\,\rho_{\hat\Pi_1}) + (1{-}w)\,\Tr(G\,\rho_{\hat\Pi_2}) \nonumber\\&
 	=\Tr(G\,[w\, \rho_{\hat\Pi_1} + (1{-}w)\,\rho_{\hat\Pi_2}]) \nonumber\\& = \Tr(G\,W_{\Pi_3}) = g(W_{\Pi_3})\,,
 	 \end{align}
where $W_{\Pi_3}$ is the usual von Neumann 
statistical operator, defined by the weighted sum of the density operators  representing the component homogeneous sub-ensembles,
 \begin{equation}\label{eq:statistical_operator}
	W_{\Pi_3}=w\,\rho_{\hat\Pi_1} + (1{-}w)\,\rho_{\hat\Pi_2}.
\end{equation}
$W_{\Pi_3}$ is clearly an element of the convex completion, $\mathcal{C}(\mathcal{P})$, of the  quantal state domain $\mathcal{P}$. Notice that, in general $\mathcal{C}(\mathcal{P}_\text{QSM})=\mathcal{C}(\mathcal{P}_\text{HG})=\mathcal{P}_\text{HG}$, and for a qubit $\mathcal{P}_\text{QSM}$ and $\mathcal{P}_\text{HG}$ map to the Bloch sphere and the Bloch ball, respectively.
An evaluation of the linear functional $g(\cdot)=\Tr(G\cdot)$  at $W_{\Pi_3}$ provides the correct expected mean value of measurements of the corresponding observable on the heterogeneous ensemble.

But $W_{\Pi_3}$ does not fully describe the heterogeneous preparation $\Pi_3$, because different preparations can yield the same linear tomography. In other words, the linear tomography---obtained by measuring the expected mean values $\overline{\langle  {q_j} \rangle}$ of  a quorum of $N^2{-}1$ independent linear observables $Q_j$ and solving $\overline{\langle  {q_j} \rangle}=\Tr(Q_j\,W)$ for $W$---is insufficient to characterize a preparation. A decomposition of $W$ into different weighted sums of  density operators has no meaning in this theory, but it emphasizes 
 that different heterogeneous preparations may result in the same linear tomography. Even within orthodox QSM ($\mathcal{P}{\,=\,} \mathcal{P}_\text{QSM} $), resolving the intrinsic quantum probabilities of homogeneous preparations from the extrinsic uncertainties of mixing requires additional independent information beyond \mbox{linear tomography.}

The measure-theoretic description of preparations enables a statistical quantum theory in which both linear and nonlinear functionals of the density operator correspond to independently and directly measurable properties of a quantum system, such as entropy, ergotropy, and adiabatic availability. A nonlinear evolution equation for the density operators in $\mathcal{P}$, i.e., for the homogeneous preparations, may provide additional nonlinear observables by measuring linear and nonlinear observables at different times. As already realized by  \mbox{\citet{park_1988_thermodynamic},} nonlinearity holds the promise of preserving and reintegrating the notion of an individual state into quantum theory. 

\section{\label{sec:lpo}Local Perception Operators (LPOs)}

In linear QM, the system's composition is specified by declaring the following: 
\begin{enumerate}
    \item The Hilbert space structure as the direct product $\Hil = \bigotimes_{\J=1}^M\Hil_\J$ of the subspaces of the $M$ component subsystems.
    \item The overall Hamiltonian operator \(H  = \sum_{\J=1}^M H_\J{\otimes} I_\Jbar + V \),  where $ H_\J$ (on $\Hil_\J$) is the local Hamiltonian of the $J$-th subsystem, $I_\Jbar$ is 
    the identity on the direct product \(\Hil_\Jbar= \bigotimes_{K\ne J}\Hil_K\) of all the other subspaces, and $V$ (on $\Hil$) is the interaction Hamiltonian.
\end{enumerate}

{The linear} 
 von Neumann law of evolution, $\dot\rho= -\imi[H,\rho]/\hbar$, has a universal structure and involves local evolutions through partial tracing, 
\begin{equation}\label{eq:partial_trace_linear}
\dot\rho_\J= -\dfrac{\imi}{\hbar}[H_\J,\rho_\J] -\dfrac{\imi}{\hbar}\Tr_\Jbar([V,\rho]).
\end{equation}
Thus, we recover the universal law $\dot\rho_\J= -\imi[H_\J,\rho_\J]/\hbar$ for the local density operator $\rho_\J=\Tr_\Jbar(\rho)$ if the subsystem $\J$ does not interact with the others (i.e., if $V=I_\J{\otimes} V_\Jbar$).

Instead, a fully nonlinear evolution equation for the density operator cannot have a universal structure because the subdivision into subsystems must be explicitly embedded into the structure of the dynamical law (see~\cite{beretta_2010_maximum} for more on this). A different subdivision requires a different equation of motion. The complex structure of the SEA evolution law reflects the cost of abandoning linearity but ensures compatibility with the crucial constraint that correlations should not build up, and signaling should not occur between subsystems, other than via the interaction Hamiltonian $V$ through the standard Schr\"{o}dinger term $-\imi[H,\rho]/\hbar$ in the evolution law.

Seldom used in composite quantum dynamics but crucial, in our opinion, are the physical observables first introduced in~\cite{beretta_1985_quantuma} and later (starting with~\cite{pelegrin_1987_colloque}) referred to as `local perceptions of global observables'.  These are represented via the `local perception operators' (LPO) on  $\Hil_\J$, defined along with their `deviation from the local mean value' operators and covariance functionals as follows:
\begin{align}
	(X)^J_\rho        & = \Tr_\Jbar[(I_\J{\otimes}\rho_\Jbar)X] \,,\label{eq:local_perception_operators}             \\
	\Delta(X)^\J_\rho & =(X)^\J_\rho-I_\J\Tr[\rho_\J(X)^\J_\rho]\,,\label{eq:deviation_X}                          \\
	(X,Y)^\J_\rho     & = \half\Tr[\rho_\J\acomm{\Delta(X)^\J_\rho}{\Delta(Y)^\J_\rho}]\,,\label{eq:covariance_XY}
\end{align}
where $\rho_\Jbar=\Tr_\J(\rho)$. For a bipartite system, $AB$, the LPOs $(X)^A_\rho$ (on $\Hil_A$) and $(X)^B_\rho$ (on $\Hil_B$) are the unique operators that, for a given $X$ on $\Hil_{AB}$ and for all states $\rho$, satisfy the identity
\begin{equation}\label{eq:local_perception_operators_id}
	\Tr[\rho_A(X)^A_\rho] =\Tr[(\rho_A{\otimes}\rho_B)X] =\Tr[\rho_B(X)^B_\rho]\,.\\
\end{equation}
This confirms that they encapsulate all the information that $A$ and $B$ can infer about the global observable X by classically sharing their local states. 

Operator $(X)^A_\rho$ can be viewed as the projection onto $\Hil_A$ of the operator $X$ weighted according to the local state $\rho_B$ of subsystem $B$. However, it is a local observable for subsystem A, which depends on the overall state, $\rho$, and the overall observable, $X$. Its local mean value, $\Tr[\rho_A(X)^A_\rho]$, differs from the mean value, $\Tr(\rho X)$, for the overall system, $AB$, except when $A$ and $B$ are uncorrelated ($\rho=\rho_A{\otimes}\rho_B$). It was dubbed `local perception' because, even if $B$ performs a local tomography and sends the measured $\rho_B$ to $A$ via classical communication, the most that $A$ can measure locally about the overall observable $X$ is $(X)^A_\rho$.

The overall energy and entropy of the composite system are locally perceived within subsystem $\J$ through the operators $(H)^\J_\rho$ and $(S)^\J_\rho$ defined on $\Hil_\J$ by Equation~(\ref{eq:local_perception_operators}), respectively, with $X=H$, the overall Hamiltonian, and $X=S(\rho)$, the overall (non-negative) entropy operator defined by
\begin{equation}\label{eq:entropy_operator}
	S(\rho)=-\Boltz{\rm{Bln}}(\rho)\,,
\end{equation}
where we define the discontinuous log function
\begin{equation}
    \Bln (x) = \begin{cases}
        \ln(x), ~~\text{for } 0< x\le 1,\\
        0, ~~\text{otherwise.}
    \end{cases}
\end{equation}
Note that the `locally perceived overall entropy' operator
\begin{equation}
	(S)^J_\rho         = -\Boltz\Tr_\Jbar[(I_\J{\otimes}\rho_\Jbar)\Bln(\rho)] \,,\label{eq:local_perception_S_operator}
\end{equation}
 is different from the `local entropy' operator
 \begin{equation}
 	S(\rho_\J)=-\Boltz\Bln (\rho_\J) \,.\label{eq:local_S_operator}
 \end{equation}
Its local mean value,  $\Tr[\rho_\J(S)^\J_\rho]=-\Boltz\Tr[(\rho_\J{\otimes}\rho_\Jbar)\Bln (\rho)]$, is different from the local entropy $\Tr[\rho_\J S(\rho_\J)]= -\Boltz\Tr[\rho_\J\ln (\rho_\J)]$. Only when $\rho=\rho_\J{\otimes}\rho_\Jbar$ are they related via
\begin{align}\label{eq:LPO_entropy}
    \begin{split}
         \Tr[\rho_\J(S)^\J_\rho]&=\Tr[\rho_\J S(\rho_\J)]+\Tr[\rho_\Jbar S(\rho_\Jbar)]\, \\
         &= -\Boltz\Tr[\rho\ln (\rho)]. 
    \end{split}
\end{align}
Likewise,  the `locally perceived overall Hamiltonian' operator $(H)^\J_\rho$ differs from the `local Hamiltonian' operator $H_\J$. Its local mean value,  $\Tr[\rho_\J(H)^\J_\rho]=\Tr[(\rho_\J{\otimes}\rho_\Jbar)H]$, is different from the local mean energy, $\Tr(\rho_\J H_\J)$, and only when $V=I_\J{\otimes} V_\Jbar$ are they related via 
\begin{equation}\label{eq:LPO_energy}
\Tr[\rho_\J(H)^\J_\rho]=\Tr(\rho_\J H_\J)+\Tr(\rho_\Jbar H_\Jbar)= \Tr(\rho H).
\end{equation}
However, it is noteworthy that, when the overall observable, $X$ is `separable for subsystem $\J$,' in the sense that  $X=X_\J{\otimes} I_\Jbar+I_\J{\otimes} X_\Jbar$, then, even if $\rho\ne\rho_\J{\otimes}\rho_\Jbar$, the deviations and covariances reduce to their local versions,
\begin{align}
	\Delta(X)^\J_\rho & =\Delta X_\J = X_\J-I_\J\Tr[\rho_\J X_\J]\,,\label{eq:deviation_separableJ}       \\
	(X,Y)^\J_\rho     & =\half\Tr[\rho_\J\acomm{\Delta X_\J}{\Delta Y_\J}]\,.\label{eq:covariance_separable}
\end{align}

In general, the mean interaction energy, $\Tr(\rho_\J V_{\J\Jbar })$, and the mutual information, $\Tr(\rho_\J \mu_{\J\Jbar })$, between subsystem $\J$ and the rest of the system, $\Jbar$, are given via the respective mean values of the following global operators:
\begin{align}
	&V_{\J\Jbar}=H-H_\J{\otimes} I_\Jbar-I_\J{\otimes} H_\Jbar\,,\label{eq:VJJbar}\\
	&\mu_{\J\Jbar}= \Bln (\rho)-\Bln (\rho_\J){\otimes} I_\Jbar-I_\J{\otimes} \Bln (\rho_\Jbar)\,,\label{eq:muJJbar}
\end{align}
whose LPOs satisfy the identities
\begin{align}
		&\Tr[\rho_\J(H)^\J_\rho]-\Tr[\rho_\J (V_{\J\Jbar})^\J_\rho ]\nonumber\\
         &\quad =\Tr(\rho_\J H_\J)+\Tr(\rho_\Jbar H_\Jbar)	\nonumber\\
         &\quad = \Tr(\rho H)-\Tr(\rho_\J V_{\J\Jbar}) \,,\label{eq:identity_VJJbar}
     \end{align}\vspace{-15pt}
\begin{align}
		&\Tr[\rho_\J(S)^\J_\rho]+\Boltz\Tr[\rho_\J (\mu_{\J\Jbar})^\J_\rho ]\nonumber\\
	&\quad =-\Boltz\Tr[\rho_\J\ln (\rho_\J)]-\Boltz\Tr[\rho_\Jbar\ln (\rho_\Jbar)]	\nonumber\\
			&\quad =-\Boltz\Tr[\rho\ln (\rho)]+ \Boltz\Tr(\rho_\J \mu_{\J\Jbar }) \,.\label{eq:identity_MI}
\end{align}

\section{\label{sec:nosignaling_lpo}No-Signaling and LPOs}

To formalize the no-signaling definition following~\cite{beretta_2005_nonlinear}, as discussed above, we adopt the view that local operations can  be acted on a subsystem, $\Jbar$, only by means of a controlled time dependence of its local Hamiltonian operator, $H_\Jbar$, or its interaction operator, $ V_{\Jbar B }$, with some other subsystem, $B$ (possibly a properly modeled environment or heat bath). Hence, we adopt the following:
\begin{Definition}[No-signaling]
If $\J$ and $\Jbar$ are non-interacting, no local aspect of the time evolution of $\J$ can be affected by local unitary operations acted on $\Jbar$, nor by other aspects of the local  time evolution of $\Jbar$.
\end{Definition}
Accordingly, consider a composite, $AB$, in state $\rho$. A local unitary operation on $B$ changes the state to
\begin{equation}\label{eq:partial_unitary_op_on_rho}
	\rho'=(I_A{\otimes} U_B)\,\rho\, (I_A{\otimes} U_B^\dagger)\,,
\end{equation}
where $U_B$ is an arbitrary unitary operator ($U_B^\dagger U_B=I_B$). Using  the properties of the partial trace, in particular ($Z_{AB}$ is a generic composite system operator),
\begin{align}\label{eq:partial_trace_properties}
	 & \Tr_B[(I_A{\otimes} X_B)Z_{AB}]=\Tr_B[Z_{AB}(I_A{\otimes} X_B)]\,,         \\
	 & \Tr_A[(I_A{\otimes} X_B)Z_{AB}(I_A{\otimes} Y_B)]= X_B\Tr_A(Z_{AB})Y_B \,,
\end{align}
we obtain the identities
\begin{align}
	\rho_B  & =\Tr_A[(I_A{\otimes} U_B^\dagger)\,\rho'\, (I_A{\otimes} U_B)]=U_B^\dagger\rho'_BU_B,
	\label{eq:no_signaling_unit_rhoB}                                                                                   \\
	\rho'_A & =\Tr_B[(I_A{\otimes} U_B)\,\rho\, (I_A{\otimes} U_B^\dagger)]\notag \\
    &=\Tr_B[(I_A{\otimes} U_B^\dagger U_B)\,\rho]\notag \\&=\Tr_B[(I_A{\otimes} I_B)\,\rho]=\rho_A, \label{eq:rhoAprime}
\end{align}
which confirms that a local operation on $B$ does not affect the local state, $\rho_A$, of $A$. This result supports the usual idea~\cite{ferrero_2004_nonlinear} that, for no-signaling, it is sufficient that the dynamical model implies evolutions of local observables that depend only on $\rho_A$. But it is seldom noted that this is not a necessary condition. 

Next, we prove that not only the local reduced state $\rho_A$ but also the LPO $(F(\rho))^A$ of any well-defined nonlinear function, $F(\rho)$, of the overall state (such as the entropy function $S(\rho)$ defined earlier) remains unaffected by local unitary operations on $B$, as per Equation~(\ref{eq:partial_unitary_op_on_rho}). Since the SEA formalism is based on these local perception operators, this result is an important lemma in the proof that SEA is no-signaling.

So, let us apply Equation~(\ref{eq:partial_unitary_op_on_rho}) to a generic observable represented by a linear operator  on $\Hil$ that we denote as $F(\rho)$, whether it is a nontrivial function of $\rho$, such as $S(\rho)$, or not a function of $\rho$, such as $H$. The corresponding LPO for subsystem A, according to the defining Equation~(\ref{eq:local_perception_operators}), is given by
\begin{equation}\label{eq:no_signaling_on_f5}
	(F(\rho))^A = \Tr_B[(I_A\otimes\rho_B)F(\rho)].
\end{equation}
A function of $\rho$ is defined from its eigen-decomposition via 
\begin{equation}
F(\rho) = \Lambda F(D)\Lambda^\dagger=\sum_jF(\lambda_j)\dyad{\lambda_j},
\end{equation} 
where  $\rho=\Lambda D\Lambda^\dagger$, $D=\sum_j\lambda_j\dyad{j}$, and $\Lambda=\sum_j\dyad{\lambda_j}{j}$.
Since unitary transformations do not alter the eigenvalues,
\begin{equation}\label{eq:no_signaling_on_f7}
	F(\rho') = \Lambda'F(D)\Lambda'^\dagger \mbox{ where } \Lambda'=(I_A\otimes U_B)\Lambda \,.
\end{equation}
Therefore, using Equation~(\ref{eq:no_signaling_unit_rhoB}) in the last step, we obtain
\begin{align}\label{eq:no_signaling_on_f8}
	 & (F(\rho'))^A = \Tr_B[(I_A\otimes\rho'_B)F(\rho')]\notag \\ &=\Tr_B[(I_A\otimes \rho'_B)\,(I_A\otimes U_B)\Lambda F(D)\Lambda^\dagger(I_A\otimes U_B^\dagger)]\notag\\ &=\Tr_B[(I_A\otimes U_B^\dagger\rho'_B U_B)\,\Lambda F(D)\Lambda^\dagger]\notag\\
	 & =\Tr_B[(I_A\otimes \rho_B)\,F(\rho)]=(F(\rho))^A\,.
\end{align}
This confirms that local unitary operations on $B$ do not affect the LPOs of $A$. Hence, the proper use of LPOs in a nonlinear evolution equation does not cause signaling issues.

\section{\label{sec:composite_SEA}\textls[-15]{The Local Structure of Dissipation in Composite-System SEA Dynamics}}

We are now ready to introduce the final but essential ingredient of a general composite-system nonlinear QM or master equation. This involves the system's structure-dependent expressions, which determine how each subsystem contributes separately to the dissipative term in the equation of motion for the overall state, $\rho$. As discussed above---and recognized in early SEA literature~\cite{beretta_1985_quantuma,beretta_2005_nonlinear,beretta_2010_maximum}---the composite-system nonlinear evolution must explicitly reflect the internal structure of the system. This requires declaring which subsystems are to be protected from nonphysical effects such as signaling,  the exchange of energy, or the build-up of correlations between non-interacting subsystems. Using the notation introduced earlier, the structure proposed in~\cite{beretta_1985_quantuma,beretta_2010_maximum} for the dissipative term, which supplements the usual Hamiltonian term, is given by 
\begin{equation}\label{eq:SEA_Composite_simp}
	\dv{\rho}{t} = -\frac{\imi}{\hbar}\comm{\mathit{H}}{\rho}-\sum_{\J=1}^M\acomm*{\mathcal{D}^\J_\rho}{\rho_\J}\otimes\rho_\Jbar\,,
\end{equation}
where the `local dissipation operators' $\mathcal{D}^\J_\rho$ (on $\Hil_\J$) may be nonlinear functions of the local observables of $\J$, the reduced state $\rho_\J$, and the local perception operators of overall observables. For the dissipative term to preserve $\Tr(\rho)$, operators $\acomm*{\mathcal{D}^\J_\rho}{\rho_\J}$ must be traceless. To preserve $\Tr(\rho H)$ [and possibly other conserved properties or charges, $\Tr(\rho C_k)$], operators  $\acomm*{\mathcal{D}^\J_\rho}{\rho_\J}(H)^\J_\rho$ [and $\acomm*{\mathcal{D}^\J_\rho}{\rho_\J}(C_k)^\J_\rho$] must also be traceless. The rate of change of the overall system entropy $s(\rho)=-\Boltz\Tr[\rho\ln(\rho)]$ is
\begin{equation}\label{eq:SEA_entropy_prod}
	\dv{s(\rho)}{t}= -\sum_{\J=1}^M\Tr[\acomm*{\mathcal{D}^\J_\rho}{\rho_\J}(S)^\J_\rho]\,.
\end{equation}
The local nonlinear evolution of subsystem $\J$ is obtained via partial tracing over $\Hil_\Jbar$, i.e., \mbox{in general,}
\begin{equation}\label{eq:SEA_local}
	\dv{\rho_\J}{t} = -\frac{\imi}{\hbar}\comm*{H_\J}{\rho_\J} -\frac{\imi}{\hbar}\Tr_\Jbar(\comm*{V}{\rho}) -\acomm*{\mathcal{D}^\J_\rho}{\rho_\J}\,,
\end{equation}
where we recall that the second term in the RHS can be expressed, for weak interactions and under well-known assumptions, in Kossakowski--Lindblad form.

Before introducing the SEA assumption, as promised after Equation~(\ref{eq:no_signaling_condition}), we note that, for all possible choices of $\mathcal{D}^\J_\rho$, Equation~(\ref{eq:SEA_Composite_simp}) defines a broad class of no-signaling nonlinear evolution equations. These form a broader class of nonlinear laws that are not restricted by the sufficient but not necessary condition that ${\rm d}\rho_\J/{\rm d}t$ be a function of $\rho_\J$ only.

Finally, one way to introduce the SEA assumption in the spirit of the fourth law of thermodynamics~\cite{beretta_2020_fourth}, i.e., to implement the maximum entropy production principle (MEPP) in the present context~\cite{beretta_2014_steepesta}, is to employ a variational principle.

\section{\label{sec:variational}General Composite System Version of the SEA Variational Principle}

For the purpose of this section, we adopt the formalism developed for this context in~\cite{beretta_1985_quantuma,gheorghiu-svirschevski_2001_nonlinear,beretta_2010_maximum}. To easily impose the constraints of preservation of non-negativity and self-adjointness of $\rho$ during its time evolution, we define the generalized square root of $\rho_\J$, $\gamma_\J(t)=\sqrt{\rho_\J(t)}\,U$ ($U$ any unitary operator), so that
\begin{equation}\label{eq:gammaDef}
	\rho_\J=\gamma_\J\gamma_\J^\dagger. 
\end{equation}
The dissipative term in Equation~(\ref{eq:SEA_local}) is 
 rewritten as
\begin{align}\label{eq:SEA_Composite_gamma}
	-\acomm*{\mathcal{D}^\J_\rho}{\rho_\J}&=\dot\gamma^d_\J\gamma_\J^\dagger+\gamma_\J\dot\gamma^{d\dagger}_\J\\ \text{ with } \dot\gamma^d_\J&=-\mathcal{D}^\J_\rho\gamma_\J \,.\label{eq:SEA_Composite_gammad}
\end{align}
Next, on the set $\mathcal{L}(\Hil_\J)$ of linear operators on $\Hil_\J$, we define the  real inner product, $( \cdot | \cdot )$,
\begin{equation}\label{eq:innerProduct}
	( X | Y)=\Tr (X^{\dagger} Y + Y^{\dagger} X )/2\,,
\end{equation}
so that the unit trace condition for $\rho_\J$ is rewritten as $( \gamma_\J | \gamma_\J)=1$, implying that the $ \gamma_\J$'s lie on the unit sphere in $\mathcal{L}(\Hil_\J)$. Their time evolutions, $\gamma_\J(t)$, follow trajectories on this sphere.
Along these trajectories, we can express the distance traveled between $t$ and $t{\,+\,}\D t$ as
\begin{equation}\label{eq:distance}
	\D\ell_\J = \sqrt{(\dot\gamma_\J|\,\hat G_\J(\gamma_\J)\,|\dot\gamma_\J)}\, \D t\,,
\end{equation}
where  $\hat G_\J(\gamma_\J)$ is some real, dimensionless, symmetric, and positive--definite operator on  $\mathcal{L}(\Hil_\J)$ (superoperator on $\Hil$) that plays the role of a local metric tensor field (and may be a nonlinear function of $\gamma_\J$).

The rates of change of the overall system entropy, $s(\rho)$, Equation~(\ref{eq:SEA_entropy_prod}), and of the overall system mean value of conserved properties $c_k(\rho)=\Tr(\rho C_k)$, where $\comm*{C_k}{H}=0$, can be rewritten as
\begin{align}
	\dv{s(\rho)}{t}&= \sum_{\J=1}^M\dot{s}|_\J \qquad \dot{s}|_\J=\Big(2(S)^\J_\rho\gamma_\J\Big|\dot\gamma^{d}_\J\Big)\,, \label{eq:SEA_entropy_prod_gamma}\\
	\dv{c_k(\rho)}{t}&= \sum_{\J=1}^M\dot{c_k}|_\J\qquad \dot{c_k}|_\J=\Big(2(C_k)^\J_\rho\gamma_\J\Big|\dot\gamma^{d}_\J\Big)\,, \label{eq:constants_gamma}
\end{align}
exhibiting additive contributions from the subsystems.

Finally, we state the variational principle that leads to expressions for the $\dot\gamma^{d}_\J$ and the $\mathcal{D}^\J_\rho$ 
 that define the composite-system version of the SEA equation of motion. The time evolution ensures that the 'direction of change' of the local trajectory $\gamma_\J(t)$, influenced by the dissipative part of the dynamics, maximizes the local contribution, $\dot{s}|_\J$, to the overall system's entropy production rate. This occurs under the constraints $\dot{c_k}|_\J=0$ that guarantee no local contribution to the rates of change of the global constants of the motion.  With the introduction of Lagrange multipliers $\vartheta_k^\J$ and $\tau_\J$ for the constraints, the $\dot\gamma^d_\J$'s are found by solving the maximization problem
\begin{equation}\label{eq:variational} 
\max_{\dot\gamma^{d}_\J}\ \Upsilon_\J= \dot{s}|_\J - \sum_k\vartheta_k^\J \dot{c_k}|_\J -\frac{\Boltz\tau_\J}{2} \Big(\dot\gamma^d_\J\Big|\,\hat G_\J\,\Big|\dot\gamma^d_\J\Big)\,,
\end{equation}
where the last constraint corresponds to the condition $(\D\ell^d_\J/\D t)^2= \text{const}$, necessary for maximizing with respect to direction only (see~\cite{beretta_2014_steepesta} for more details). Taking the variational derivative of $\Upsilon_\J$ with respect to $|\dot\gamma^{d}_\J)$ and setting it equal to zero, we obtain
\begin{equation}\label{eq:Lagrangian} 
	\frac{\delta\Upsilon_\J}{|\delta\dot\gamma^{d}_\J)}= \Big|2(M)^\J_\rho\gamma_\J\Big) -\Boltz\tau_\J\, \hat G_\J\,\Big|\dot\gamma^d_\J\Big)=0\,,
\end{equation}
where we used the identity $(X|\,\hat G_\J=\hat G_\J\,|X) $, following from the symmetry of $\hat G_\J$. We define (following~\cite{beretta_2009_nonlinear,beretta_2019_time}) the 'locally perceived  non-equilibrium Massieu operator'\clearpage
\begin{equation}\label{eq:Massieu_operator} 
	(M)^\J_\rho=(S)^\J_\rho-\sum_k\vartheta_k^\J(C_k)^\J_\rho\,.
\end{equation}
Equation~(\ref{eq:Lagrangian}) yields
\begin{equation}\label{eq:SEA_general_supermetric} 
	\Big|\dot\gamma^d_\J\Big)=\frac{1}{\Boltz\tau_\J}\hat G_\J^{-1} 	\Big|2(M)^\J_\rho\gamma_\J\Big)\,,
\end{equation}
where the Lagrange multipliers $\vartheta_k^\J$ (implicit in $(M)^\J_\rho$) are the solution of the system of equations, obtained by substituting Equation~(\ref{eq:SEA_general_supermetric}) into the conservation constraints,
\begin{equation}\label{eq:multiplierssuper} 
	\Big((C_\ell)^\J_\rho\gamma_\J\Big|\hat G_\J^{-1}\Big|(M)^\J_\rho\gamma_\J\Big)=0\quad \forall \ell\,.
\end{equation}
This system can be solved explicitly for the $\vartheta_k^\J$'s using Cramer's rule, to obtain convenient expressions for the $\dot\gamma^{d}_\J$'s as ratios of determinants (as in the original formulations). The $\vartheta_k^\J$'s are nonlinear functionals of $\rho$ that may be interpreted as 'local non-equilibrium entropic potentials' conjugated with the conserved properties. For example, for $C_2=H$, the Lagrange multiplier $\vartheta_2^\J$ plays the role of 'local non-equilibrium inverse temperature' conjugated with the locally perceived energy, and for the stable equilibrium states of the SEA dynamics, it coincides with the  thermodynamic inverse temperature $\Boltz\beta_\J$ (see below).

Similarly to what was achieved 
in~\cite{beretta_2014_steepesta} for a non-composite system, we  define the 'local  non-equilibrium affinity' operators
 \begin{equation}\label{eq:noneq_affinity} 
 	|\Lambda_\J)= \hat G_\J^{-1/2}\Big|2(M)^\J_\rho\gamma_\J\Big)\,,
 \end{equation}
so that the overall rate of entropy production becomes
\begin{equation}\label{eq:entropy_production_SEA_super} 
	\dv{s(\rho)}{t}= \sum_{\J=1}^M\frac{(\Lambda_\J|\Lambda_\J)}{\Boltz\tau_\J}\,.
\end{equation}
$(\Lambda_\J|\Lambda_\J)$ is the norm of $2(M)^\J_\rho\gamma_\J$ with respect to the metric $\hat G_\J^{-1}$ and may be interpreted as the 'degree of disequilibrium' of subsystem $\J$. Hence, the necessary and sufficient condition for the overall state to be locally non-dissipative (no contribution to the overall entropy production from subsystem $\J$) is that operator $2(M)^\J_\rho\gamma_\J$ vanishes.

However, in order for the equation of motion (\ref{eq:SEA_local}) to result independent of the unitary operators $U$ used (in  $\gamma_\J=\sqrt{\rho_\J}\,U$) to define the generalized square roots of $\rho_\J$, 
 we further restrict the choice of the metric superoperator $\hat G_\J$. We assume that $\hat G_\J= L_\J^{-1} \hat I_\J$, with $L_\J$ being some strictly positive, hermitian operator on $\Hil_\J$, possibly a nonlinear function of $\rho_\J$, so that $\hat G_\J|X) = |L_\J^{-1} X)$, $\hat G^{-1}_J|X) = |L_\J X)$, 
\begin{equation} 
	(X\gamma_\J|\hat G^{-1}_J|Y\gamma_\J) = \half\Tr\left[ \rho_\J(X^\dagger L_\J Y+Y^\dagger L_\J X)\right].
\end{equation}
With the recollection of Equations~(\ref{eq:SEA_Composite_gamma}) and (\ref{eq:SEA_Composite_gammad}), and using Equation~(\ref{eq:SEA_general_supermetric}), the dissipative term in Equation~(\ref{eq:SEA_local}) becomes 
\begin{equation}\label{eq:SEA_general} 
	-\acomm*{\mathcal{D}^\J_\rho}{\rho_\J}=\frac{2}{\Boltz\tau_\J}\left[L_\J (M)^\J_\rho\rho_\J +\rho_\J (M)^\J_\rho L_\J\right]\,,
\end{equation}
and the system of equations that determines the Lagrange multipliers  $\vartheta_k^\J$ in $(M)^\J_\rho$ is
\begin{equation}\label{eq:multipliers} 
\Tr\left(\rho_\J\Big[(C_\ell)^\J_\rho L_\J(M)^\J_\rho+(M)^\J_\rho L_\J(C_\ell)^\J_\rho\Big]\right)=0\quad \forall \ell
\end{equation}
so that the  dependence on  $\gamma_\J$ is only through the product  $\gamma_\J \gamma_\J^\dagger$, i.e., the local state \mbox{operator $\rho_\J$.}

The metric operator $L_\J^{-1}/\Boltz\tau_\J$ plays a role analogous to the symmetric thermal conductivity tensor $\hat k$ in heat transfer theory. In that context, \(\hat k\) defines the general near-equilibrium linear relationship, $|q'')=-\hat k\,|\nabla T)$, between the heat flux vector $q''$ and the  conjugated 'degree of disequilibrium' vector, i.e., the temperature gradient $\nabla T$. Here, Equation~(\ref{eq:SEA_general}) expresses a more general linear relationship between the local evolution operator $\dot\gamma^d_\J$ and the non-equilibrium Massieu operator, $(S)^\J_\rho - \sum_k\vartheta_k^\J(C_k)^\J_\rho$. This relation is more general, as it holds not only near equilibrium but also anywhere far from equilibrium. In the present quantum modeling context,  it represents the nonlinear SEA extension into the far-non-equilibrium domain of Onsager’s linear near-equilibrium theory, with reciprocity naturally embedded through the symmetry of any metric. For example, if $L_\J$ commutes with the local Hamiltonian $H_\J$, its eigenvalues can assign different relaxation times to the local energy levels, capturing their uneven contributions to the rate of local energy redistribution described by the SEA dissipator $\mathcal{D}^\J_\rho$.

Like in heat transfer, the conductivity tensor for an isotropic material is $\hat k= k\,\hat I$; here, an analogous simplification is obtained when $L_\J=  I_\J$, the identity operator on $\Hil_\J$. This corresponds to assuming a uniform Fisher--Rao metric.

Another noteworthy observation is that operators $C_k$, to be global constants of the motion (or 'charges'), must commute with the global Hamiltonian operator $H$, but they need not commute with each other. Therefore, the SEA formalism may also find application in the framework of quantum thermodynamic resource theories  that contemplate 'non-commuting charges' and 'non-Abelian thermal states', as discussed in \mbox{\citet{YungerHalpern_naturecomm_2016} and \citet{YungerHalpern_PRL_2023}.}

\section{\label{sec:SEAsimple}Simplest Composite-System SEA Equation of Motion}

For simplicity, we proceed by assuming the following: (1) a uniform Fisher--Rao metric, with $\hat G_\J=\hat I_\J$, so that $\hat G_\J|X)=|X)$ and $L_\J=I_\J$; and (2) only two global conserved properties, $c_1(\rho)=\Tr(\rho I)$ (the normalization condition) and $c_2(\rho)=\Tr(\rho H)$ (mean energy conservation). 

The first assumption allows the obtaining of
\begin{equation}\label{eq:DJ_Def}
    \mathcal{D}^\J_\rho =-\frac{2}{\Boltz\tau_\J}\left[(S)^\J_\rho- {\textstyle \sum_k}\vartheta^\J_k(C_k)^\J_\rho\right].
\end{equation}
and the system of Equation~(\ref{eq:multipliers}) reduces to
\begin{align}\label{eq:betakJ_system}
	 & \sum_k\vartheta^\J_k\Tr[\rho_\J\acomm{(C_\ell)^\J_\rho}{(C_k)^\J_\rho}]\nonumber \\
	 &= \Tr[\rho_\J\acomm{(S)^\J_\rho}{(C_\ell)^\J_\rho}]\, .
\end{align}

The second assumption, $C_1=I$ and $C_2=H$, together with definitions (\ref{eq:deviation_X}) and (\ref{eq:covariance_XY}), allows the writing of the SEA dissipators in the compact forms
\begin{align}\label{eq:DJ_compact}
&-\acomm*{\mathcal{D}^J_\rho}{\rho_J} =\dfrac{2}{\Boltz\tau_\J}\acomm*{\Delta(M)^\J_\rho}{\rho_J} \nonumber\\ &=\dfrac{2}{\Boltz\tau_\J}\tfrac{\vmqty{\acomm*{\Delta(S)^\J_\rho}{\rho_J} & \acomm*{\Delta (H)^\J_\rho}{\rho_J} \\
			(H,S)^\J_\rho & 	(H,H)^\J_\rho }}{\displaystyle	(H,H)^\J_\rho }\,,
\end{align}
and the local non-equilibrium inverse temperatures conjugated with the locally perceived energy as
\begin{equation}\label{eq:DJ_Def1}
	\vartheta_2^\J =\frac{	(H,S)^\J_\rho}{(H,H)^\J_\rho}\,.
\end{equation}

As a result, the local evolution of each subsystem, $\J$, is along the direction of steepest ascent of the locally perceived overall composite-system entropy, compatible with the conservation of the locally perceived overall composite-system energy and the conditions of unit-trace and non-negativity of the overall density operator, $\rho$. The rate of entropy production may be expressed as
\begin{align}\label{eq:S_production}
	\dv{s(\rho)}{t}&=\sum_{J=1}^M  \dfrac{4}{\Boltz\tau_\J}(M,M)^\J_\rho \nonumber\\
	&=\sum_{J=1}^M  \dfrac{4}{\Boltz\tau_\J}\tfrac{\vmqty{(S,S)^\J_\rho & 	(H,S)^\J_\rho \\
			(H,S)^\J_\rho & 	(H,H)^\J_\rho }}{\displaystyle	(H,H)^\J_\rho } \,,
\end{align}
showing clearly that it is non-negative since the numerators in the summation are Gram determinants. When the $\J$-th term in the sum vanishes, we say that the local state $\rho_\J$ is 'non-dissipative'. If all local states are non-dissipative, then so is the overall state $\rho$. A non-dissipative $\rho$ represents an equilibrium state if it commutes with $H$; otherwise, it belongs to a unitary limit cycle of the dynamics. 

For $\rho$ to be non-dissipative, there must exist $\beta_J$s, 
 such that, for every $\J$,
\begin{equation}\label{eq:rhoJnondiss1}
	\rho_\J\Delta(\Bln (\rho))^\J_\rho=-\beta_J\rho_\J\Delta (H)^\J_\rho\,,
\end{equation}
or, equivalently,
\begin{equation}\label{eq:rhoJnondiss}
	\rho_\J\Delta\left[\Bln (\rho_\J){+}\beta_J H_\J{+}\beta_J(V_{\J\Jbar })^\J_\rho{+}(\mu_{\J\Jbar })^\J_\rho\right]=0\,,
\end{equation}
where $V_{\J\Jbar }$ is the Hamiltonian interaction operator defined in Equation~(\ref{eq:VJJbar}), and $\mu_{\J\Jbar}$ the mutual information operator defined in Equation~(\ref{eq:muJJbar}). Clearly, for $V_{\J\Jbar }=0$ and $\mu_{\J\Jbar}=0$, Equation~(\ref{eq:rhoJnondiss}) is satisfied by the Gibbs states
\begin{equation}\label{eq:rhoGibbs}
	\rho_\J=\exp(-\beta_\J H_\J)/\Tr[\exp(-\beta_\J H_\J)]\,.
\end{equation}

Section \ref{sec:example103} provides a numerical illustration of local time evolutions that converge to limit cycles obeying Equation~(\ref{eq:rhoJnondiss}). These examples also illustrate another noteworthy possibility. Specifically, without violating the no-signaling condition nor the second-law principle of global entropy non-decrease---which is always guaranteed by virtue of \mbox{Equation~(\ref{eq:S_production})} and the properties of Gram determinants---it is possible that the local entropy of a subsystem be decreasing during part of the time evolution. Indeed, for correlated states, the non-decrease in local entropy is not a second-law requirement. We hope that the observation that a local entropy decrease may be thermodynamically consistent in the presence of strong entanglement will stimulate the design of experimental verifications, as well as foundational discussions in the framework of quantum thermodynamic resource theories, even beyond and outside the SEA framework. 

Regarding no-signaling, we note the following:
\begin{enumerate}
    \item If subsystem $\J$ is non-interacting, $V_{\J\Jbar }=0$, then 
    \begin{equation}
    \Delta (H)^\J_\rho= H_\J-I_\J\Tr(\rho_\J H_\J)=\Delta H_\J
    \end{equation}
    and 
    \begin{equation}
    (H,H)^\J_\rho=\Tr[\rho_J(\Delta H_\J)^2]
    \end{equation}
    depend only on the local $H_\J$ and $\rho_\J$;
    \item If $\J$ is uncorrelated, $\mu_{\J\Jbar}=0$, then 
    \begin{equation}
    \Delta (\Bln (\rho))^\J_\rho=  \Bln (\rho_\J)-I_\J\Tr(\rho_\J  \ln(\rho_\J))
    \end{equation}
    and 
    \begin{equation}
    (\Bln (\rho),\Bln (\rho))^\J_\rho=\Tr[\rho_\J\left(\ln(\rho_\J)\right)^2]
    \end{equation}
    depend only on the local  $\rho_\J$.    
\end{enumerate}

{Therefore, it is} 
 only when $\J$ is both non-interacting and uncorrelated in that 
 its local dissipation operator, $\mathcal{D}^\J_\rho$, depends only on the local $H_\J$ and $\rho_\J$. In this case, the local equation of motion Equation~(\ref{eq:SEA_local}), along with $\mathcal{D}^\J_\rho$ given by Equation~(\ref{eq:DJ_compact}), reduces exactly to the non-composite system version of SEA evolution~\cite{beretta_1984_quantuma}. Instead, if $\J$ is either interacting or correlated, $\mathcal{D}^\J_\rho$ and, therefore, the local nonlinear SEA evolution, according to Equations~(\ref{eq:SEA_local}) and (\ref{eq:DJ_compact}), is determined not only by the local $H_\J$ and $\rho_\J$ but also by the local perceptions of the overall Hamiltonian operator $H$ and/or the overall entropy operator $\Bln (\rho)$, nonetheless without violating the no-signaling condition.

To prove no-signaling, assume that subsystem $\J$ is correlated but not interacting with any of the subsystems in $\Jbar$. Now, switch on an arbitrary interaction that may involve the subsystems in $\Jbar$ but not subsystem $\J$, so that the only change in the overall Hamiltonian $H$ is the term  $ H_\Jbar$, which changes to $ H'_\Jbar$. Within $\J$, the locally perceived deviation operators are $\Delta (H)^\J_\rho$ and $\Delta (\Bln (\rho))^\J_\rho$, and hence, the SEA dissipator $\mathcal{D}^\J_\rho$, as well as all the terms in the RHS of Equation~(\ref{eq:SEA_local}), are not modified by such a change. Therefore, acting within $\Jbar$ makes it impossible to affect the time evolution of $\rho_\J$ and any local observable of $\J$.

We also mention the following in the spirit of non-linear evolutions. In the case of a general open quantum dynamics, the requirement that a map, $\mathcal{W}_t(\rho)$, be completely positive and trace-preserving (CPTP) is quite restrictive~\cite{simmons_1981_completely, raggio_1982_remarks, simmons_1982_another}. In fact, the reduced dynamics of the subsystem in interaction with the environment need not be CPTP~\cite{pechukas_1994_reduced, alicki_1995_comment, pechukas_1995_pechukas}. Here, we define CPTP as follows. If $\mathcal{M}(\rho_s)$ is a map acting on a subsystem in state $\rho_s$, which interacts with an initially uncorrelated environment in state $\rho_E$ and is positive and trace-preserving---meaning it preserves the trace and ensures that the evolved matrix remains semi-definite (maintaining probability conservation)---then $\mathcal{M}(\rho_s)$ is CPTP if and only if the extended map, $\Lambda: \mathcal{M}(\rho_s){\otimes} \mathit{I}_N$, remains positive for all $N$s~\cite{simmons_1981_completely}. As has been argued in the literature, the  CPTP condition is restrictive because, along with the underlying assumption of Markovianity, it requires a preparation of the initial state into a product state of system and environment. But, if there are strong initial correlations between system and environment, or in fact, if the evolution is non-Markovian in nature, then this requirement fails, and we are forced to consider PTP (positive and trace-preserving) maps only~\cite{rivas_2010_entanglement, jagadish_2019_measurea}. Given that we are dealing with a theory that does not restrict itself to Markovian evolution, and that there is no imposition of weak interaction between the subsystems, SEA being PTP satisfies the requirements for a suitable nonlinear model of quantum thermodynamics \mbox{prohibiting signaling.}\clearpage

\section{\label{sec:sea_example}Non-Interacting Qubits}

In extreme cases, Equation~(\ref{eq:rhoJnondiss}) shows that, even if the subsystems are entangled, and therefore, the local states $\rho_\J$ are mixed, operators $\mathcal{D}^\J_\rho$ may vanish. Equations~(\ref{eq:SEA_Composite_simp}) and (\ref{eq:SEA_local}) reduce to the standard Schr\"{o}dinger equation, and the trajectory in state space is a limit cycle of the SEA dynamics. A noteworthy particular case is when the overall system is in a pure state. Then, $\Bln (\rho)=0$ and standard unitary evolutions of pure states emerge as limit cycles of the nonlinear SEA dynamics. In this section, we discuss a few less trivial examples, to numerically illustrate some general features of Equation~(\ref{eq:SEA_Composite_simp}) with $\mathcal{D}^\J_\rho$ given by Equation~(\ref{eq:DJ_compact}).

We consider examples of mixed and correlated initial states of a two-qubit composite, $AB$, that belong to the special class
\begin{align}\label{eq:general_mixed_state}
			\begin{split}
&\rho =\quarter\Big[\mathrm{I}_4+\!\!\!\!\!\!\sum_{j=\{x,y,z\}}\!\!\!\!(a_j\,\sigma_j{\otimes}\mathrm{I}_2 +b_j\,\mathrm{I}_2{\otimes}\sigma_j+c_j\,\sigma_j{\otimes}\sigma_j)\Big]  \\
&= \quarter\!\!\left(\!\begin{smallmatrix}1{+}a_z{+}b_z{+}c_z\!\!\!\!\!   &b_x{-}\imi b_y       & a_x{-}\imi a_y      & c_x{-}c_y\\
		b_x{+}\imi b_y       &\!\!\!\!\!1{+}a_z{-}b_z{-}c_z\!\!\!\!\!   &c_x{+}c_y &a_x{-}\imi a_y\\
		a_x{+}\imi a_y       &c_x{+}c_y &\!\!\!\!\!1{-}a_z{+}b_z{-}c_z\!\!\!\!\!   &b_x{-}\imi b_y\\
		c_x{-}c_y &a_x{+}\imi a_y       & b_x{+}\imi b_y      & \!\!\!\!\!1{-}a_z{-}b_z{+}c_z  \end{smallmatrix}\right),
	 		\end{split}
\end{align}
with real $a_j$'s, $b_j$'s, $c_j$'s, such that {$a^2=\bm{a}{\cdot}\bm{a}\le 1$,} 
 $b^2=\bm{b}{\cdot}\bm{b}\le 1$, $c^2=\bm{c}{\cdot}\bm{c}\le 3- a^2-b^2$, plus other conditions necessary for non-negativity (see e.g.,~\cite{horodecki_1995_Bell,bruning_2012_parametrizations,byrd_2003_characterization}. We will denote the eigenvalues of $\rho$ as $\lambda_j$ and, for shorthand, define
\begin{equation}
	\eta_j=-\Boltz\Bln (\lambda_j)\,. \label{eq:surprise}
\end{equation}
We further assume that the two qubits are non-interacting and have local Hamiltonian operators given by $H_\J = \omega_\J\bm{h}^\J\cdot\bm{\sigma}_\J$, where $\bm{\sigma}_\J$ denotes the 3-vector formed by the local Pauli operators of subsystem $\J$, and $\bm{h}^\J$ is the local Hamiltonian unit 3-vector, so that
\begin{equation}\label{eq:BDS_hamilton_total}
		\begin{split}
&H = H_A{\otimes}\mathit{I}_2+\mathit{I}_2{\otimes} H_B \\
& \hphantom{H}	=\omega_A\bm{h}^A{\cdot}\bm{\sigma}_A {\otimes}\mathit{I}_2+\mathit{I}_2{\otimes}  \omega_B\bm{h}^B{\cdot}\bm{\sigma}_B\,,\\
&(H)_\rho^A=H_A{+}\mathit{I}_2\,\omega_B\bm{h}^B{\cdot}\bm{b}\,,\\  &(H)_\rho^B=\mathit{I}_2\,\omega_A\bm{h}^A{\cdot}\bm{a}{+}H_B\\
&\Delta(H)^A = \Delta H_A\,, \quad \Delta(H)^B = \Delta H_B\,,\\
&(H,H)_\rho^A = [1-(\bm{h}^A{\cdot}\bm{a})^2]\, \omega^2_A\,,\\ 
&(H,H)_\rho^B = [1-(\bm{h}^B{\cdot}\bm{b})^2]\,\omega^2_B\,,
		\end{split}
\end{equation}

\subsection{\label{sec:Bell_diagonal}Bell Diagonal States}

Equation~(\ref{eq:general_mixed_state}) gives Bell diagonal states~\cite{horodecki_1996_informationtheoretic, lang_2010_quantum} (BDS) if $a_j=b_j=0$ for all $j$s (and Werner states~\cite{werner_1989_quantum} if, in addition, $c_j=4w/3-1$ for all $j$s),
 \begin{align}\label{eq:bell_states}
 		\begin{split}
 	&\rho^\Bell=\quarter\left(\begin{smallmatrix}1{+}c_z   &0       & 0      & c_x{-}c_y\\
 		0       &1{-}c_z   &c_x{+}c_y &0\\
 		0       &c_x{+}c_y &1{-}c_z   &0\\
 		c_x{-}c_y &0       & 0      & 1{+}c_z  \end{smallmatrix}\right) =\half\left(\begin{smallmatrix}\lambda_2{+}\lambda_3   &0       & 0      & \lambda_3{-}\lambda_2\\
 			0       &\lambda_1{+}\lambda_4   &\lambda_4{-}\lambda_1 &0\\
 			0       &\lambda_4{-}\lambda_1 &\lambda_1{+}\lambda_4   &0\\
 			\lambda_3{-}\lambda_2 &0       & 0      & \lambda_2{+}\lambda_3 \end{smallmatrix}\right),\\
 		&\ \ =	\rho_A{\otimes}\rho_B+ \sum_{j=\{x,y,z\}}c_j\,\sigma_j{\otimes}\sigma_j\,
 		\end{split}
 \end{align}
whose eigen-decomposition may be written as
\begin{align}
	\begin{split}
		&\rho^\Bell   = U^\Bell\text{diag}(\lambda_1,\lambda_2,\lambda_3,\lambda_4) (U^\Bell)^\dagger,  \\
		&U^\Bell  =  \frac{\sigma_x{\otimes} \mathrm{I}_2-\sigma_z{\otimes}\sigma_x}{\sqrt{2}}={\textstyle \frac{1}{\sqrt{2}} }\text{\footnotesize$\begin{pmatrix}0  &-1 & 1 &0\\
		-1 &0  &0  &1\\
		1  &0  &0  &1\\
		0  &1  & 1 &0\end{pmatrix}$}\,, \\
		&\lambda_1   = \quarter(1{-}c_x{-}c_y{-}c_z)\,,\quad
		\lambda_2  =  \quarter(1{-}c_x{+}c_y{+}c_z)\,, \\
	&	\lambda_3  = \quarter(1{+}c_x{-}c_y{+}c_z)\,,\quad 
		\lambda_4  =  \quarter(1{+}c_x{+}c_y{-}c_z)\,.
	\end{split}
\end{align}
The overall entropy operator, Equation~(\ref{eq:entropy_operator}), becomes 
\begin{align}
	\begin{split}
&	S(\rho^\Bell)  = U^\Bell\,[\text{diag}(\eta_1,\eta_2,\eta_3,\eta_4)] (U^\Bell)^\dagger  \\
	 	&\qquad=\half\text{\footnotesize$\begin{pmatrix}\eta_2{+}\eta_3   &0       & 0      & \eta_3{-}\eta_2\\
			0       &\eta_1{+}\eta_4   &\eta_4{-}\eta_1 &0\\
			0       &\eta_4{-}\eta_1 &\eta_1{+}\eta_4   &0\\
			\eta_3{-}\eta_2 &0       & 0      & \eta_2{+}\eta_3 \end{pmatrix}$},\\
   \end{split}
\end{align}
and the local operators of the SEA formalism
\begin{align}
\begin{split}  &  \rho_A =\rho_B=\half I_2\,,\\
  &  (\Bln(\rho))^A =(\Bln(\rho))^B=-{\textstyle\frac{q}{4\Boltz}}\mathrm{I}_2 \,,\\
    &     \acomm*{(\Bln(\rho))^A}{\rho_A} =\acomm*{(\Bln(\rho))^B}{\rho_B}=-{\textstyle\frac{q}{4\Boltz}}\mathrm{I}_2 \,,\\
  &     \acomm*{\Delta(\Bln(\rho))^A}{\rho_A} =\acomm*{\Delta(\Bln(\rho))^B}{\rho_B}=0 \,,\\
  &(H,H)_\rho^A =  \omega^2_A\,,\quad  (H,H)_\rho^B =\omega^2_B\,, \\
  &(H,\Bln(\rho))_\rho^A =  (H,\Bln(\rho))_\rho^B =0\,. \\
\end{split}
\end{align}
where $q$ is defined as
\begin{equation}
	q=\eta_1+ \eta_2 +\eta_3 + \eta_4\,.
\end{equation} 
Therefore, somewhat surprisingly, we find that
\begin{equation}
-\acomm*{\mathcal{D}^A_\rho}{\rho_A}=-\acomm*{\mathcal{D}^B_\rho}{\rho_B}=0\,,
\end{equation}
i.e., the Bell diagonal states are non-dissipative limit cycles of the nonlinear SEA dynamics under any Hamiltonian. But most neighboring and other states in the class defined by Equation~(\ref{eq:general_mixed_state}) are dissipative, as we see in the following examples.

\subsection{\label{sec:example102}Separable, but Correlated Mixed States}
For a simple example of correlated but separable mixed states, assume Equation~(\ref{eq:general_mixed_state}) with $a_x=a$, $b_z=b$, and $a_y=a_z=b_x=b_y=c_x=c_y=c_z=0$, so that
\begin{align}\label{eq:initial_state_10_2}
	\begin{split}
		\rho&=\quarter\left(\!\begin{smallmatrix}1{+}b  &0 & a &0\\
				0 &1{-}b  &0  &a\\
				a  &0  &1{+}b  &0\\
				0  &a  & 0 &1{-}b  \end{smallmatrix}\right) =	\rho_A{\otimes}\rho_B-\quarter ab\,\sigma_x{\otimes}\sigma_z\,,\\
		\lambda_1  & = \quarter ( 1-a-b)\,,\quad
		\lambda_2  = \quarter ( 1-a+b)\,, \\
		\lambda_3 &= \quarter ( 1+a-b)\,, \quad
		\lambda_4  = \quarter ( 1+a+b).
	\end{split}
\end{align}
If the two non-interacting qubits have local Hamiltonians $H_A=\sigma_z$ and $H_B=\sigma_x$, we find
\begin{align}\label{eq:Dissipation_1}
	-\acomm*{\mathcal{D}^A_\rho}{\rho_A} & = \dfrac{(1-a^2)(bf-g)}{2\Boltz\tau_A}\,\sigma_x, \\
	-\acomm*{\mathcal{D}^B_\rho}{\rho_B} & = \dfrac{(1-b^2)(af-p)}{2\Boltz\tau_B}\,\sigma_z,
\end{align}
where $f$, $g$, and $p$ are defined as
\begin{align}\label{eq:mixed_parameters}
    \begin{split}
        f&=\eta_1- \eta_2 -\eta_3 + \eta_4,\\
        g&=\eta_1 + \eta_2 - \eta_3 - \eta_4,\\
        p&=\eta_1 - \eta_2 + \eta_3 - \eta_4,         
    \end{split}
\end{align}
so that  the nonlinear evolution is clearly nontrivial. However, it preserves the zero mean energies of both qubits, and while the overall entropy increases and mutual information partially fades away, it drives the overall state towards a non-dissipative correlated state with maximally mixed marginals. We proved above that signaling is impossible, even though $\mathcal{D}^A_\rho$ depends not only on $a$ but also on $b$, and $\mathcal{D}^B_\rho$ on $a$, which agrees with our no-signaling condition in Equation~(\ref{eq:no_signaling_condition}).

\begin{figure}[H] 
	\includegraphics[width=0.85\linewidth]{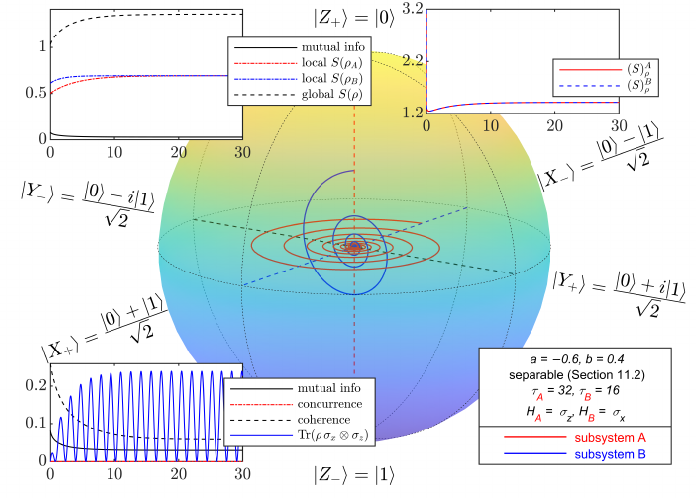}
	\caption{{Bloch-ball representation} 
  of the time evolutions of the local density operators $\rho_A$ and $\rho_B$ of non-interacting qubits $A$ and $ B$ with local Hamiltonians $H_A=\sigma_z$ and $H_B=\sigma_x$. The evolution follows the steepest locally perceived entropy ascent equation, Equation~(\ref{eq:SEA_Composite_simp}), with $\hbar=1$ and $\Boltz=1$. The initial state of the composite system $AB$ is the separable, correlated mixed state given by Equation~(\ref{eq:initial_state_10_2}) with  $a=-0.6$ and $b=0.4$. Insets show the time evolutions of global [$S(\rho)$], local [$S(\rho_A)$, $S(\rho_B)$], and locally perceived [$(S)^A_\rho=(S)^B_\rho$] entropies, mutual information [$S(\rho_A)+S(\rho_B)-S(\rho)$], \mbox{concurrence---a} coherence measure [$\Tr(\rho^2H^2-\rho H\rho H)$], and the Bell observable $\Tr(\rho\,\sigma_x{\otimes}\sigma_z)$. The local evolutions approach the respective local Gibbs states. Without violating the no-signaling condition, dissipation causes a (non-complete) reduction in mutual information and coherence, while  the Bell observable builds up to a steady state oscillation.}
	\label{fig:Bloch102}
\end{figure}

Figure \ref{fig:Bloch102} shows a Bloch-ball representation of the time evolutions of the local density operators $\rho_A$ and $\rho_B$ for an initial state in the class considered in this section, Equation~(\ref{eq:initial_state_10_2}), with $a=-0.6$ and $b=0.4$. The time evolution is computed through a numerical integration of  Equation~(\ref{eq:SEA_Composite_simp}).   The local evolutions of the two non-interacting qubits approach asymptotically the respective local Gibbs states, i.e., the states of maximum local entropy for the given initial mean local energies, for which Equation~(\ref{eq:rhoJnondiss}) reduce to
\begin{align}\label{eq:Gibbs_states_non_iinteracting}
	\begin{split}
		\Delta(\Bln (\rho_A))&=-\beta_A\,\Delta H_A,\\
		\Delta(\Bln (\rho_B))&=-\beta_B\,\Delta H_B.
	\end{split}
\end{align} 
 Mutual information, $S(\rho_A)+S(\rho_B)-S(\rho)$, and a measure of coherence, $\Tr(\rho^2H^2-\rho H\rho H)$ $=\half\Tr(K^\dagger K)$, where $K=i[H,\rho]$, are reduced through the SEA dissipation terms but do not vanish. Indeed, the  Bell observable, $\Tr(\rho\,\sigma_x{\otimes}\sigma_z)$, which is initially zero, builds up to a periodic oscillation of constant amplitude.

\subsection{\label{sec:example103}Entangled, Separable, and 
	Correlated Mixed States}

\begin{figure}[H] 
 	\includegraphics[height=0.58\columnwidth]{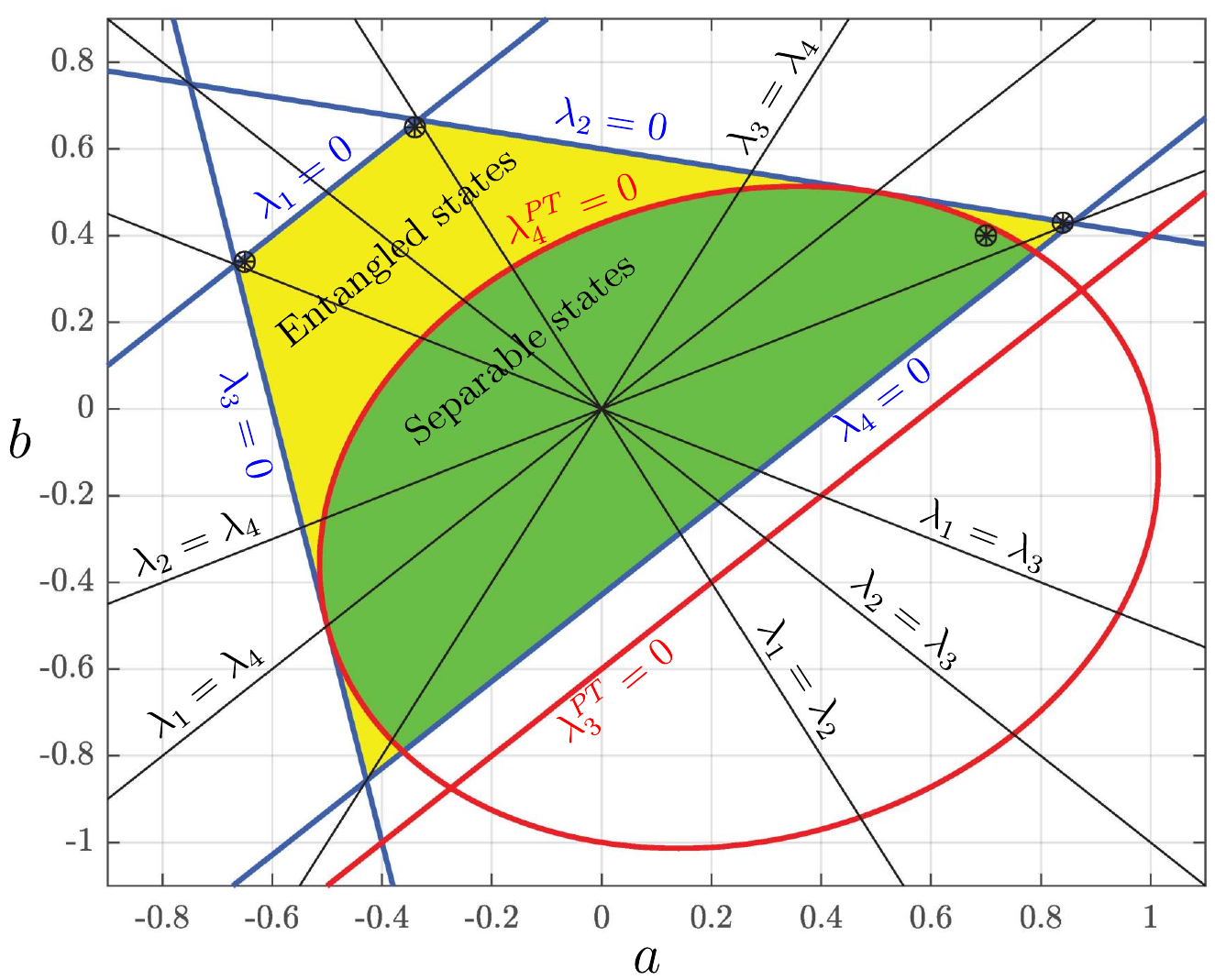}
 	\caption{\textls[-5]{\hl{Ranges of admissible} 
 values of parameters $a$ and $b$ for the separable and entangled initial states in the class defined in Section~\ref{sec:example103}. The four points denoted with the symbol $\mathrlap{\rm o}{\rm *}$ represent the initial states chosen to illustrate the local time evolutions in Figures~\ref{fig:Bloch103_1}--\ref{fig:Bloch105}. The two points near the $\lambda_1=0$ line represent the strongly entangled states for which Figures~\ref{fig:Bloch104} and \ref{fig:Bloch105} show a local entropy decrease for one of the subsystems, while the second-law principle of global entropy non-decrease is not violated.}}
 	\label{fig:possibleValues}
\end{figure}

For a slightly more elaborate example that includes entangled mixed states, assume  Equation~(\ref{eq:general_mixed_state}) with $a_x=a_z=a/\sqrt{2}$, $b_x=b_z=b/\sqrt{2}$, and $c_x=c_y=c_z=2(a-b)/3$,  so that  the eigenvalues of $\rho$ and those of its partial transpose are
\begin{align}\label{eq:initial_state_10_3}
	\begin{split}
		\lambda_1  = \quarter (1+a-b)\,,\quad & \lambda_2  = \twelth ( 3-a-5b)\,,  \\
		\lambda_3  =\twelth ( 3+5a+b)\,,\quad & \lambda_4  =  \twelth (3-7a+7b)\,, 
	\end{split}
\end{align}
and
\begin{align}
	\begin{split}
	\lambda^{PT}_1  & = \twelth (3+a-b)\,, \quad   \lambda^{PT}_2   =  \twelth (3-5a+5b)\,,  \\
	\lambda^{PT}_3  &= \twelth (3+2a-2b+\sqrt{d})\,, \\
        \lambda^{PT}_4  &=  \twelth (3+2a-2b-\sqrt{d})\,, 
	\end{split}
\end{align}
with $d=25a^2 - 14ab + 25b^2$. Figure~\ref{fig:possibleValues} shows the complete set of admissible pairs of values of parameters $a$ and $b$ for the set of correlated mixed states considered in this example, which encompasses both separable and entangled states. For instance, considering $a=-b$ ($\lambda_2=\lambda_3$), these states are separable for $-3/14\le b\le 1/4$ and entangled for $1/4<b\le 1/2$. 
We compute explicitly each term in Equation~(\ref{eq:DJ_compact}), also for this class of states, to find
 \begin{align}
 \begin{split}
  &\acomm*{(S)^A}{\rho_A} =-\dfrac{(1{-}a^2)(f {-} 5b p)}{20\sqrt{2}} (\sigma_x{+}\sigma_z)\,,\\
  &\acomm*{(S)^B}{\rho_B} =\dfrac{(1{-}b^2)(g {+} 5a p)}{20\sqrt{2}} (\sigma_x{+}\sigma_z)\,,
 \end{split}
\end{align}\vspace{-12pt}
\begin{align}
\begin{split}
	  &(H,S)^A_\rho = -\dfrac{(1{-}a^2)(f {-} 5b p)\omega_A}{20\sqrt{2}} (h_x^A{+}h_z^A) \,,\\
   &(H,S)^B_\rho = \dfrac{(1{-}b^2)(g {+} 5a p)\omega_B}{20\sqrt{2}} (h_x^B{+}h_z^B) \,,
\end{split}
\end{align}\vspace{-12pt}
\begin{align}
\begin{split}
	    &(H,H)^A_\rho = [1{-}  \half (h_x^A{+}h_z^A)^2a^2]\omega^2_A \,,\\
   &(H,H)^B_\rho = [1{-}  \half (h_x^B{+}h_z^B)^2b^2]\omega^2_B \,,
\end{split}
\end{align}\vspace{-12pt}
\begin{align}
\begin{split}
	   &	-\acomm*{\mathcal{D}^A_\rho}{\rho_A}  =- \dfrac{(1 {-}a^2)(f {-}5bp)}{5\sqrt{2}[2{-}(h_x^A{+}h_z^A)^2a^2]\Boltz\tau_A}\,\zeta_A\,,  \\
  & -\acomm*{\mathcal{D}^B_\rho}{\rho_B}  = \dfrac{(1 {-}b^2)(g{+}5ap)}{5\sqrt{2}[2{-}(h_x^B{+}h_z^B)^2b^2]\Boltz\tau_B}\,\zeta_B\,.
\end{split}
\end{align}
Here, $f$, $g$, and $p$ are defined as\vspace{-6pt}

 \begin{align}
 	\begin{split}
 		f&=3\eta_1 - 5\eta_2 +5 \eta_3 -3\eta_4\,,\\
 		g&=3\eta_1 +5\eta_2 - 5\eta_3 - 3\eta_4\,,\\
 		p&=\eta_1 - \eta_2 - \eta_3 + \eta_4\,,    
   \end{split}
\end{align}\vspace{-12pt}
\begin{align}
\begin{split}
	 		\zeta_A&=\Big[1- \half (h_x^A{+}h_z^A)^2a^2+\half (h_x^A{+}h_z^A)a^2\Big](\sigma_x{+}\sigma_z)\\
 		&-  (h_x^A{+}h_z^A)  \bm{h}^A{\cdot}\bm{\sigma}_A  \,,\\
 		\zeta_B&=\Big[1- \half (h_x^B{+}h_z^B)^2b^2+\half (h_x^B{+}h_z^B)b^2\Big](\sigma_x{+}\sigma_z)\\
 		&-  (h_x^B{+}h_z^B)  \bm{h}^B{\cdot}\bm{\sigma_B}  \,.
\end{split}
\end{align}
For example, if the two non-interacting qubits $A$ and $B$ have local Hamiltonians $H_A=\sigma_z$ and $H_B=\sigma_x$, we find
\begin{align}\label{eq:Dissipation_2}
	-\acomm*{\mathcal{D}^A_\rho}{\rho_A} & =- \dfrac{(1 {-}a^2)(f {-}5bp)}{5\sqrt{2}(2{-}a^2)\Boltz\tau_A}\,\sigma_x,  \\
	-\acomm*{\mathcal{D}^B_\rho}{\rho_B} & = \dfrac{(1 {-}b^2)(g{+}5ap)}{5\sqrt{2}(2{-}b^2)\Boltz\tau_B}\,\sigma_z,
\end{align}
so that again the nonlinear evolution is clearly nontrivial in the sense that the local nonlinear evolution of $A$ ($B$) does not depend only on $\rho_A$ ($\rho_B$), despite being no-signaling.

\begin{figure}[H] 
	\includegraphics[width=0.85\linewidth]{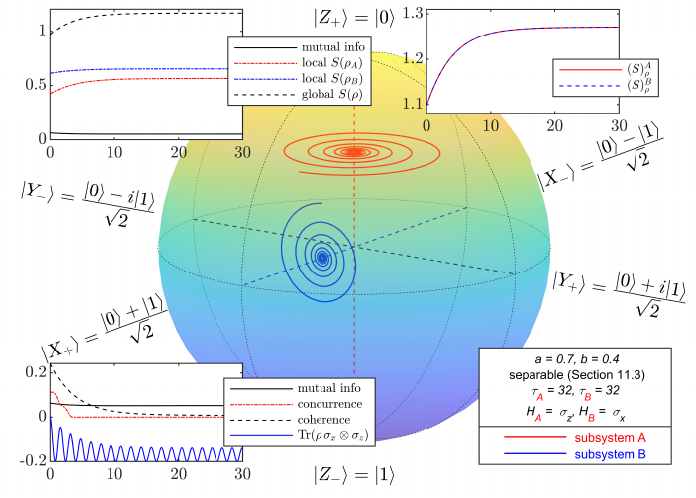}
	\caption{\hl{Bloch-ball representation} of the time evolutions of the local density operators $\rho_A$ and $\rho_B$ of non-interacting qubits $A$ and $ B$ with local Hamiltonians $H_A=\sigma_z$ and $H_B=\sigma_x$, resulting from numerical integration of the steepest locally perceived entropy ascent equation of motion, Equation~(\ref{eq:SEA_Composite_simp}). The initial state of the composite system $AB$ is the separable, correlated mixed state in the class of states defined in Section~\ref{sec:example103} with $a=0.7$ and $b=0.4$. The local evolutions approach the respective local Gibbs states. Without violating the no-signaling condition, dissipation causes a (non-complete) reduction in mutual information and coherence, while the Bell observable converges to a steady state oscillation. Insets are similar to 
		Figure~\ref{fig:Bloch102}.}
\label{fig:Bloch103_1}
\end{figure}

\begin{figure}[H]
        \includegraphics[width=0.85\linewidth]{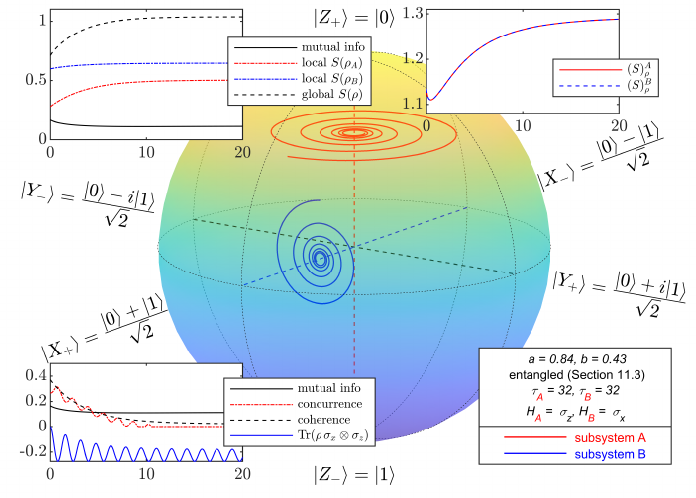}
    \caption{\hl{Bloch-ball representations} of the time evolution of two non-interacting qubits, obtained from numerical integration of Equation~(\ref{eq:SEA_Composite_simp}) with initial entangled mixed states from the class defined in Section~\ref{sec:example103}. The values of \((a,b)\) are \((0.84,0.43)\). Due to entanglement, local evolutions approach limit cycles with local entropy lower than the corresponding local Gibbs states.
    Insets are similar to 
     Figure~\ref{fig:Bloch102}.}
    \label{fig:Bloch103}
\end{figure}

\begin{figure}[H]
        \includegraphics[width=0.85\linewidth]{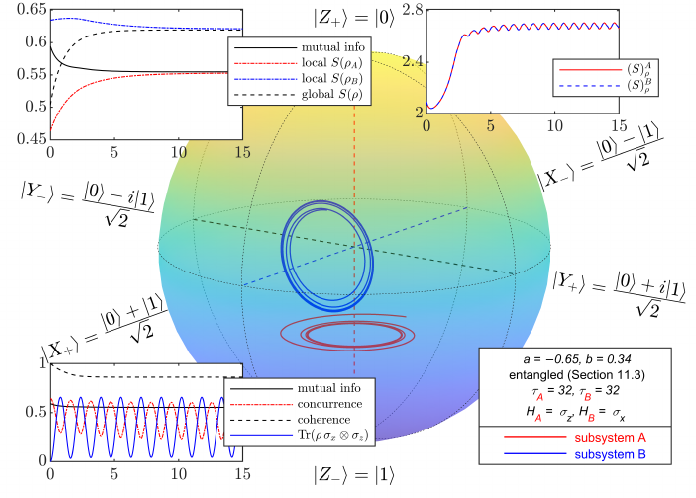}
    \caption{Bloch-ball representations of the time evolution of two non-interacting qubits, obtained from numerical integration of Equation~(\ref{eq:SEA_Composite_simp}) with initial entangled mixed states from the class defined in Section \ref{sec:example103}. The values of \((a,b)\) are \(({-}0.65,0.34)\). Due to entanglement, local evolutions approach limit cycles with local entropy significantly lower than the corresponding local Gibbs states. Strong entanglement leads to a local entropy decrease in subsystem $B$ and $A$, respectively, without violating the global entropy non-decrease dictated by the second law.
    Insets are presented 
     in Figure~\ref{fig:Bloch102}.}
    \label{fig:Bloch104}
\end{figure}

\begin{figure}[H]
        \includegraphics[width=0.85\linewidth]{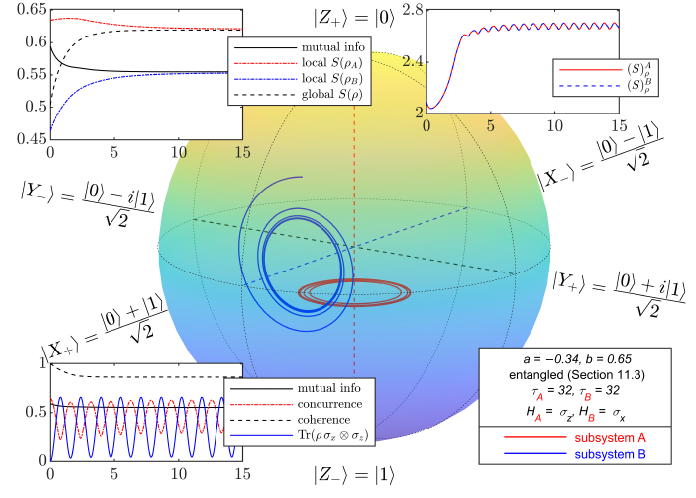}
    \caption{Bloch-ball representations of the time evolution of two non-interacting qubits, obtained from numerical integration of Equation~(\ref{eq:SEA_Composite_simp}) with initial entangled mixed states from the class defined in Section~\ref{sec:example103}. The values of \((a,b)\) are \(({-}0.34,0.65)\). Due to entanglement, local evolutions approach limit cycles with local entropy significantly lower than the corresponding local Gibbs states. Strong entanglement leads to a local entropy decrease in subsystem $B$ and $A$, respectively, without violating the global entropy non-decrease dictated by the second law.
    Insets are similar to 
    Figure~\ref{fig:Bloch102}.}
    \label{fig:Bloch105}
\end{figure}

For initial states in the class considered in this section, with parameter values $a$ and $b$ corresponding to the four points in Figure~\ref{fig:possibleValues}, Figures~\ref{fig:Bloch103_1}--\ref{fig:Bloch105} depict the typical time evolution of the local density operators $\rho_A$ and $\rho_B$ in the local Bloch-balls. These results are obtained from the numerical integration of the steepest locally perceived entropy ascent equation of motion, Equation~\eqref{eq:SEA_Composite_simp}, for non-interacting subsystems $A$ and $B$.

\vspace{-8pt}

\vspace{-6pt}

\vspace{-6pt}

Whereas, for a separable initial state, Figure~\ref{fig:Bloch103_1} shows that the local states approach the respective local Gibbs states, it is clear from Figures~\ref{fig:Bloch103}--\ref{fig:Bloch105} that the presence of entanglement does not allow the local states to approach the respective local Gibbs states.  Indeed, the local evolutions approach non-dissipative unitary limit cycles where $\acommInline{\mathcal{D}^A}{\rho_A}  = 0$ and $\acommInline{\mathcal{D}^B}{\rho_B}  = 0$, and therefore, Equation~(\ref{eq:rhoJnondiss}) is satisfied. Without violating the no-signaling condition, dissipation causes a (non-complete) reduction in mutual information and  coherence, while the Bell observable $\Tr(\rho\,\sigma_x{\otimes}\sigma_z)$ and concurrence converge to steady state oscillations.

\section{\label{sec:conclusion}Conclusions}

In this work, we have explored the no-signaling principle and its implications for nonlinear dynamics in quantum theory. We examined the measure-theoretic representation of mixed ensembles by providing a consistent framework for handling nonlinearities and describe ensemble mixing unambiguously. This approach also offers a conceptual basis for developing nonlinear quantum dynamical models for non-equilibrium systems and addressing unresolved questions about individual quantum states. Additionally, we reviewed the original philosophical motivations underlying the steepest entropy ascent (SEA) nonlinear evolution law, which trades linearity for strong thermodynamic consistency.

Our analysis demonstrates that the SEA formalism provides a valid framework for describing evolution or designing master equations. These can be applied in nonlinear extensions of quantum mechanics (QM) or in phenomenological models of open quantum systems within quantum thermodynamics. Our key focus was the no-signaling principle. We introduced a criterion for no-signaling based on local perception operators (LPOs), a generalization of the traditional density-operator-based definition. Unlike the conventional assumption that local evolution depends solely on local properties, SEA formalism incorporates the local effects of preexisting correlations and coherence through LPOs. Our detailed examination of the foundations and properties of LPOs highlights their potential as key tools for constructing signaling-free, nonlinear, and non-local non-equilibrium dynamical models, even outside of the SEA framework.

By inherently respecting the no-signaling condition, the SEA formalism upholds the principle of causality and aligns with the axiomatic paradigm proposed by \citet{popescu_1998_causalitya} for quantum theories. Furthermore, the SEA approach addresses conceptual challenges such as the Schr\"{o}dinger--Park paradox, offering insights into a thermodynamically consistent integration of quantum mechanics and non-equilibrium thermodynamics. Related to the no-signaling property, our discussion of LPOs and supporting numerical examples strengthens the conjecture---supported by prior numerical results from \citet{cano-andrade_2015_steepestentropyascent}---that SEA evolution is universally compatible with decoherence principles. Specifically, SEA evolution ensures that correlations and entanglement cannot build up between non-interacting systems but can decay partially or completely, a process mediated via LPOs.

An intriguing implication of SEA formalism is its compatibility with thermodynamic principles while allowing for scenarios where, under strong entanglement, the local entropy of a subsystem may temporarily decrease during the evolution, without violating the no-signaling condition or the global entropy non-decrease dictated by the second law. We emphasize that, differently from the Lindbladian `bottom-up' approach in which quantum thermodynamic second law(s) emerge from statistical approximations of the linear unitary dynamics induced by system-environment interactions, the SEA `top-down' approach builds strong second-law compatibility directly into the nonequilibrium dynamics variational principle, constrained by the phenomenological details about the system's structure, subsystems' interactions, and environmental effects. However, because of the inherent nonlinearity of the SEA formalism, the solution becomes more complex, and a full analytical calculation can become intractable~\cite{ray_2022_steepestb}. Additionally, the relaxation times $\tau_\J$ remain heuristic parameters, which are not derived from first principles and need to be tuned with respect to a given experimental set-up.

We hope this study provides a foundational basis for exploring applications of SEA modeling in quantum computing. Our comprehensive approach and observations aim to stimulate constructive discussions and inspire further studies, particularly in developing thermodynamically consistent models for non-equilibrium processes. These findings have potential applications in modeling the dynamics of quantum systems, advancing quantum technologies, and contributing to the framework of quantum thermodynamic resource theories, even beyond the SEA paradigm.

\vspace{6pt}
\authorcontributions{R.K.R.---idea. R.K.R. and G.P.B. contributed equally to the computation and writing of the manuscript. All authors have read and agreed to the published version of the~manuscript.}

\funding{From the Institute for Basic Science (IBS) in the Republic of Korea through Project No. IBS-R024-D1}

\acknowledgments{R.K.R. is grateful to {Alok Pan} 
 of the Indian Institute of Technology Hyderabad for many useful discussions and to the Wolfram Publication Watch Team for providing full access to online Mathematica~\cite{wolframresearchinc._2022_mathematica}.
G.P.B. dedicates this paper to the memory of  James L. Park, with  gratitude for his crystal-clear, profound, and pioneering contributions to quantum theory, and in particular, for the deeply inspiring seminar he delivered at MIT in April~1979.}

\conflictsofinterest{The authors declare that they have no known competing financial interests or
personal relationships that could have appeared to influence the work reported in this paper.}

\begin{adjustwidth}{-\extralength}{0cm}
\reftitle{References}

\begin{thebibliography}{999}

\end{thebibliography}


\begin{thebibliography}{999}

\bibitem[Kosloff(2019)]{kosloff_2019_quantum}
Kosloff, R.
\newblock Quantum Thermodynamics and Open-Systems Modeling.
\newblock {\em  J. Chem. Phys.} {\bf 2019}, {\em 150},~204105.
\newblock {\url{https://doi.org/10.1063/1.5096173}}.

\bibitem[Chru{\'s}ci{\'n}ski(2022)]{chruscinski_2022_dynamicalb}
Chru{\'s}ci{\'n}ski, D.
\newblock Dynamical Maps beyond {{Markovian}} Regime.
\newblock {\em Phys. Rep.} {\bf 2022}, {\em 992},~1--85.
\newblock {\url{https://doi.org/10.1016/j.physrep.2022.09.003}}.

\bibitem[Schmidt and Gemmer(2023)]{schmidt_2023_stochastic}
Schmidt, H.J.; Gemmer, J.
\newblock Stochastic {{Thermodynamics}} of a {{Finite Quantum System Coupled}}
  to {{Two Heat Baths}}.
\newblock {\em Entropy} {\bf 2023}, {\em 25},~504.
\newblock {\url{https://doi.org/10.3390/e25030504}}.

\bibitem[{Riera-Campeny} et~al.(2021){Riera-Campeny}, Sanpera, and
  Strasberg]{riera-campeny_2021_quantum}
{Riera-Campeny}, A.; Sanpera, A.; Strasberg, P.
\newblock Quantum {{Systems Correlated}} with a {{Finite Bath}}:
  {{Nonequilibrium Dynamics}} and {{Thermodynamics}}.
\newblock {\em PRX Quantum} {\bf 2021}, {\em 2},~010340.
\newblock {\url{https://doi.org/10.1103/PRXQuantum.2.010340}}.

\bibitem[Manzano et~al.(2021)Manzano, Subero, Maillet, Fazio, Pekola, and
  Rold{\'a}n]{manzano_2021_thermodynamics}
Manzano, G.; Subero, D.; Maillet, O.; Fazio, R.; Pekola, J.P.; Rold{\'a}n,
  {\'E}.
\newblock Thermodynamics of {{Gambling Demons}}.
\newblock {\em Phys. Rev. Lett.} {\bf 2021}, {\em 126},~080603.
\newblock {\url{https://doi.org/10.1103/PhysRevLett.126.080603}}.

\bibitem[Simmons and Park(1981)]{simmons_1981_essentiala}
Simmons, R.F.; Park, J.L.
\newblock The Essential Nonlinearity {{ofN-level}} Quantum Thermodynamics.
\newblock {\em Found. Phys.} {\bf 1981}, {\em 11},~297--305.
\newblock {\url{https://doi.org/10.1007/BF00726270}}.

\bibitem[Beretta(1986)]{beretta_1986_theorema}
Beretta, G.P.
\newblock A Theorem on {{Lyapunov}} Stability for Dynamical Systems and a
  Conjecture on a Property of Entropy.
\newblock {\em J. Math. Phys.} {\bf 1986}, {\em 27},~305--308.
\newblock {\url{https://doi.org/10.1063/1.527390}}.

\bibitem[Polkovnikov et~al.(2023)Polkovnikov, Gramolin, Kaplan, Rajendran, and
  Sushkov]{polkovnikov_2023_experimental}
Polkovnikov, M.; Gramolin, A.V.; Kaplan, D.E.; Rajendran, S.; Sushkov, A.O.
\newblock Experimental {{Limit}} on {{Nonlinear State-Dependent Terms}} in
  {{Quantum Theory}}.
\newblock {\em Phys. Rev. Lett.} {\bf 2023}, {\em 130},~040202.
\newblock {\url{https://doi.org/10.1103/PhysRevLett.130.040202}}.

\bibitem[Melnychuk et~al.(2024)Melnychuk, Giaccone, Bornman, Cervantes,
  Grassellino, Harnik, Kaplan, Nahal, Pilipenko, Posen, Rajendran, and
  Sushkov]{melnychuk_2024_improved}
Melnychuk, O.; Giaccone, B.; Bornman, N.; Cervantes, R.; Grassellino, A.;
  Harnik, R.; Kaplan, D.E.; Nahal, G.; Pilipenko, R.; Posen, S.;  et~al.
\newblock An {{Improved Bound}} on {{Nonlinear Quantum Mechanics}} Using a
  {{Cryogenic Radio Frequency Experiment}}. \emph{arXiv} \textbf{2024}, arXiv:quant-ph/2411.09611. {\url{https://doi.org/10.48550/arXiv.2411.09611}}.

\bibitem[Somhorst et~al.(2023)Somhorst, {van der Meer}, Correa~Anguita,
  Schadow, Snijders, {de Goede}, Kassenberg, Venderbosch, Taballione, Epping,
  {van den Vlekkert}, Timmerhuis, Bulmer, Lugani, Walmsley, Pinkse, Eisert,
  Walk, and Renema]{somhorst_2023_quantum}
Somhorst, F.H.B.; {van der Meer}, R.; Correa~Anguita, M.; Schadow, R.;
  Snijders, H.J.; {de Goede}, M.; Kassenberg, B.; Venderbosch, P.; Taballione,
  C.; Epping, J.P.;  et~al.
\newblock Quantum Simulation of Thermodynamics in an Integrated Quantum
  Photonic Processor.
\newblock {\em Nat. Commun.} {\bf 2023}, {\em 14},~3895.
\newblock {\url{https://doi.org/10.1038/s41467-023-38413-9}}.

\bibitem[Yunger~Halpern et~al.(2016)Yunger~Halpern, Faist, Oppenheim, and
  Winter]{YungerHalpern_naturecomm_2016}
Yunger~Halpern, N.; Faist, P.; Oppenheim, J.; Winter, A.
\newblock Microcanonical and resource-theoretic derivations of the thermal
  state of a quantum system with noncommuting charges.
\newblock {\em Nat. Commun.} {\bf 2016}, {\em 7},~1--7.
\newblock {\url{https://doi.org/10.1038/ncomms12051}}.

\bibitem[Hatsopoulos and
  Gyftopoulos(1976{\natexlab{a}})]{hatsopoulos_1976_unified}
Hatsopoulos, G.N.; Gyftopoulos, E.P.
\newblock A Unified Quantum Theory of Mechanics and Thermodynamics. {{Part I}}.
  {{Postulates}}.
\newblock {\em Found. Phys.} {\bf 1976}, {\em 6},~15--31.
\newblock {\url{https://doi.org/10.1007/BF00708660}}.

\bibitem[Hatsopoulos and
  Gyftopoulos(1976{\natexlab{b}})]{hatsopoulos_1976_unifieda}
Hatsopoulos, G.N.; Gyftopoulos, E.P.
\newblock A Unified Quantum Theory of Mechanics and Thermodynamics. {{Part
  IIa}}. {{Available}} Energy.
\newblock {\em Found. Phys.} {\bf 1976}, {\em 6},~127--141.
\newblock {\url{https://doi.org/10.1007/BF00708955}}.

\bibitem[Hatsopoulos and
  Gyftopoulos(1976{\natexlab{c}})]{hatsopoulos_1976_unifiedb}
Hatsopoulos, G.N.; Gyftopoulos, E.P.
\newblock A Unified Quantum Theory of Mechanics and Thermodynamics. {{Part
  IIb}}. {{Stable}} Equilibrium States.
\newblock {\em Found. Phys.} {\bf 1976}, {\em 6},~439--455.
\newblock {\url{https://doi.org/10.1007/BF00715033}}.

\bibitem[Hatsopoulos and
  Gyftopoulos(1976{\natexlab{d}})]{hatsopoulos_1976_unifiedc}
Hatsopoulos, G.N.; Gyftopoulos, E.P.
\newblock A Unified Quantum Theory of Mechanics and Thermodynamics. {{Part
  III}}. {{Irreducible}} Quantal Dispersions.
\newblock {\em Found. Phys.} {\bf 1976}, {\em 6},~561--570.
\newblock {\url{https://doi.org/10.1007/BF00715108}}.

\bibitem[Beretta et~al.(1984)Beretta, Gyftopoulos, Park, and
  Hatsopoulos]{beretta_1984_quantuma}
Beretta, G.P.; Gyftopoulos, E.P.; Park, J.L.; Hatsopoulos, G.N.
\newblock Quantum Thermodynamics. {{A}} New Equation of Motion for a Single
  Constituent of Matter.
\newblock {\em Il Nuovo Cimento B} {\bf 1984}, {\em 82},~169--191.
\newblock {\url{https://doi.org/10.1007/BF02732871}}.

\bibitem[Beretta et~al.(1985)Beretta, Gyftopoulos, and
  Park]{beretta_1985_quantuma}
Beretta, G.P.; Gyftopoulos, E.P.; Park, J.L.
\newblock Quantum Thermodynamics. {{A}} New Equation of Motion for a General
  Quantum System.
\newblock {\em Il Nuovo Cimento B} {\bf 1985}, {\em 87},~77--97.
\newblock {\url{https://doi.org/10.1007/BF02729244}}.

\bibitem[Maddox(1985)]{maddox_1985_uniting}
Maddox, J.
\newblock Uniting Mechanics and Statistics.
\newblock {\em Nature} {\bf 1985}, {\em 316},~11.
\newblock {\url{https://doi.org/10.1038/316011a0}}.

\bibitem[Beretta(1987)]{pelegrin_1987_colloque}
Beretta, G.P.
\newblock {The role of stability in the unification of mechanics and
  thermodynamics}.
\newblock In \emph{Colloque International sur la Stabilité:
  Recueil des Conférences}; P{\'e}legrin, M., Ed.; ANAE-ONERA-CERT, Toulouse, France, 
 1987; pp. 87--104.
\newblock {\url{https://hal.science/hal-04050717} (accessed on 19 September 2025)}.

\bibitem[{Gheorghiu-Svirschevski}(2001{\natexlab{a}})]{gheorghiu-svirschevski_2001_nonlinear}
{Gheorghiu-Svirschevski}, S.
\newblock Nonlinear Quantum Evolution with Maximal Entropy Production.
\newblock {\em Phys. Rev. A} {\bf 2001}, {\em 63},~022105.
\newblock {\url{https://doi.org/10.1103/PhysRevA.63.022105}}.

\bibitem[{Gheorghiu-Svirschevski}(2001{\natexlab{b}})]{gheorghiu-svirschevski_2001_addendum}
{Gheorghiu-Svirschevski}, S.
\newblock Addendum to ``{{Nonlinear}} Quantum Evolution with Maximal Entropy
  Production''.
\newblock {\em Phys. Rev. A} {\bf 2001}, {\em 63},~054102.
\newblock {\url{https://doi.org/10.1103/PhysRevA.63.054102}}.

\bibitem[Beretta(2005)]{beretta_2005_nonlinear}
Beretta, G.P.
\newblock {{Nonlinear Extensions of Schr{\"o}dinger}}--{{von Neumann Quantum
  Dynamics}}: {{A Set of Necessary Conditions for Compatibitliy with
  Thermodynamics}}.
\newblock {\em Mod. Phys. Lett. A} {\bf 2005}, {\em 20},~977--984.
\newblock {\url{https://doi.org/10.1142/S0217732305017263}}.

\bibitem[Beretta(2006)]{beretta_2006_nonlinearb}
Beretta, G.P.
\newblock Nonlinear Model Dynamics for Closed-System, Constrained,
  Maximal-Entropy-Generation Relaxation by Energy Redistribution.
\newblock {\em Phys. Rev. E} {\bf 2006}, {\em 73},~026113.
\newblock {\url{https://doi.org/10.1103/PhysRevE.73.026113}}.

\bibitem[Beretta(2009)]{beretta_2009_nonlinear}
Beretta, G.P.
\newblock Nonlinear Quantum Evolution Equations to Model Irreversible Adiabatic
  Relaxation with Maximal Entropy Production and Other Nonunitary Processes.
\newblock {\em Rep. Math. Phys.} {\bf 2009}, {\em 64},~139--168.
\newblock {\url{https://doi.org/10.1016/S0034-4877(09)90024-6}}.

\bibitem[Beretta(2010)]{beretta_2010_maximum}
Beretta, G.P.
\newblock Maximum Entropy Production Rate in Quantum Thermodynamics.
\newblock {\em J. Phys. Conf. Ser.} {\bf 2010}, {\em
  237},~012004.
\newblock {\url{https://doi.org/10.1088/1742-6596/237/1/012004}}.

\bibitem[{Cano-Andrade} et~al.(2015){Cano-Andrade}, Beretta, and {von
  Spakovsky}]{cano-andrade_2015_steepestentropyascent}
{Cano-Andrade}, S.; Beretta, G.P.; {von Spakovsky}, M.R.
\newblock Steepest-Entropy-Ascent Quantum Thermodynamic Modeling of Decoherence
  in Two Different Microscopic Composite Systems.
\newblock {\em Phys. Rev. A} {\bf 2015}, {\em 91},~013848.
\newblock {\url{https://doi.org/10.1103/PhysRevA.91.013848}}.

\bibitem[Von~Spakovsky and Gemmer(2014)]{vonspakovsky_2014_trends}
Von~Spakovsky, M.R.; Gemmer, J.
\newblock Some Trends in Quantum Thermodynamics.
\newblock {\em Entropy} {\bf 2014}, {\em 16},~3434--3470.
\newblock {\url{https://doi.org/10.3390/e16063434}}.

\bibitem[Ray(2022)]{ray_2022_steepestb}
Ray, R.K.
\newblock Steepest Entropy Ascent Solution for a Continuous-Time Quantum
  Walker.
\newblock {\em Phys. Rev. E} {\bf 2022}, {\em 106},~024115.
\newblock {\url{https://doi.org/10.1103/PhysRevE.106.024115}}.

\bibitem[Tabakin(2017)]{tabakin_2017_modela}
Tabakin, F.
\newblock Model Dynamics for Quantum Computing.
\newblock {\em Ann. Phys.} {\bf 2017}, {\em 383},~33--78.
\newblock {\url{https://doi.org/10.1016/j.aop.2017.04.013}}.

\bibitem[Tabakin(2023)]{tabakin_2023_locala}
Tabakin, F.
\newblock Local Model Dynamics for Two Qubits.
\newblock {\em Ann. Phys.} {\bf 2023}, {\em 457},~169408.
\newblock {\url{https://doi.org/10.1016/j.aop.2023.169408}}.

\bibitem[Park(1970)]{park_1970_concept}
Park, J.L.
\newblock The Concept of Transition in Quantum Mechanics.
\newblock {\em Found. Phys.} {\bf 1970}, {\em 1},~23--33.
\newblock {\url{https://doi.org/10.1007/BF00708652}}.

\bibitem[Ghirardi(2013)]{ghirardi_2013_entanglement}
Ghirardi, G.
\newblock Entanglement, {{Nonlocality}}, {{Superluminal Signaling}} and
{{Cloning}}.
\newblock In \emph{Advances in {Quantum Mechanics}}; Bracken, P., Ed.; IntechOpen, London, UK, 2013; pp. 565--594.
\newblock {\url{https://doi.org/10.5772/56429}}

\bibitem[Wootters and Zurek(1982)]{wootters_1982_single}
Wootters, W.K.; Zurek, W.H.
\newblock A Single Quantum Cannot Be Cloned.
\newblock {\em Nature} {\bf 1982}, {\em 299},~802--803.
\newblock {\url{https://doi.org/10.1038/299802a0}}.

\bibitem[Dieks(1982)]{dieks_1982_communication}
Dieks, D.
\newblock Communication by {{EPR}} Devices.
\newblock {\em Phys. Lett. A} {\bf 1982}, {\em 92},~271--272.
\newblock {\url{https://doi.org/10.1016/0375-9601(82)90084-6}}.

\bibitem[Eberhard(1978)]{eberhard_1978_bells}
Eberhard, P.H.
\newblock Bell's Theorem and the Different Concepts of Locality.
\newblock {\em Nuovo Cimento B} {\bf 1978}, {\em 46},~392--419.
\newblock {\url{https://doi.org/10.1007/BF02728628}}.

\bibitem[Simon et~al.(2001)Simon, Bu{\v z}ek, and
  Gisin]{simon_2001_nosignaling}
Simon, C.; Bu{\v z}ek, V.; Gisin, N.
\newblock No-{{Signaling Condition}} and {{Quantum Dynamics}}.
\newblock {\em Phys. Rev. Lett.} {\bf 2001}, {\em 87},~170405.
\newblock {\url{https://doi.org/10.1103/PhysRevLett.87.170405}}.

\bibitem[B{\'o}na(2003)]{bona_2003_comment}
B{\'o}na, P.
\newblock Comment on ``{{No-Signaling Condition}} and {{Quantum Dynamics}}''.
\newblock {\em Phys. Rev. Lett.} {\bf 2003}, {\em 90},~208901.
\newblock {\url{https://doi.org/10.1103/PhysRevLett.90.208901}}.

\bibitem[Simon et~al.(2003)Simon, Bu{\v z}ek, and Gisin]{simon_2003_simon}
Simon, C.; Bu{\v z}ek, V.; Gisin, N.
\newblock Simon, Bužek, and Gisin Reply:.
\newblock {\em Phys. Rev. Lett.} {\bf 2003}, {\em 90},~208902.
\newblock {\url{https://doi.org/10.1103/PhysRevLett.90.208902}}.

\bibitem[Weinberg(1989)]{weinberg_1989_testing}
Weinberg, S.
\newblock Testing Quantum Mechanics.
\newblock {\em Ann. Phys.} {\bf 1989}, {\em 194},~336--386.
\newblock {\url{https://doi.org/10.1016/0003-4916(89)90276-5}}.

\bibitem[Gisin(1990)]{gisin_1990_weinberg}
Gisin, N.
\newblock Weinberg's Non-Linear Quantum Mechanics and Supraluminal
  Communications.
\newblock {\em Phys. Lett. A} {\bf 1990}, {\em 143},~1--2.
\newblock {\url{https://doi.org/10.1016/0375-9601(90)90786-N}}.

\bibitem[Polchinski(1991)]{polchinski_1991_weinberg}
Polchinski, J.
\newblock Weinberg's Nonlinear Quantum Mechanics and the
  {{Einstein-Podolsky-Rosen}} Paradox.
\newblock {\em Phys. Rev. Lett.} {\bf 1991}, {\em 66},~397--400.
\newblock {\url{https://doi.org/10.1103/PhysRevLett.66.397}}.

\bibitem[W{\'o}dkiewicz and Scully(1990)]{wodkiewicz_1990_weinberg}
W{\'o}dkiewicz, K.; Scully, M.O.
\newblock Weinberg's Nonlinear Wave Mechanics.
\newblock {\em Phys. Rev. A} {\bf 1990}, {\em 42},~5111--5116.
\newblock {\url{https://doi.org/10.1103/PhysRevA.42.5111}}.

\bibitem[Czachor(1996)]{czachor_1996_nonlinear}
Czachor, M.
\newblock Nonlinear {{Schr}}ödinger Equation and Two-Level
  Atoms.
\newblock {\em Phys. Rev. A} {\bf 1996}, {\em 53},~1310--1315.
\newblock {\url{https://doi.org/10.1103/PhysRevA.53.1310}}.

\bibitem[Czachor(1991)]{czachor_1991_mobility}
Czachor, M.
\newblock Mobility and Non-Separability.
\newblock {\em Found. Phys. Lett.} {\bf 1991}, {\em 4},~351--361.
\newblock {\url{https://doi.org/10.1007/BF00665894}}.

\bibitem[Gisin and Rigo(1995)]{gisin_1995_relevant}
Gisin, N.; Rigo, M.
\newblock Relevant and Irrelevant Nonlinear {{Schrodinger}} Equations.
\newblock {\em J. Phys. A Math. Gen.} {\bf 1995}, {\em
  28},~7375--7390.
\newblock {\url{https://doi.org/10.1088/0305-4470/28/24/030}}.

\bibitem[Abrams and Lloyd(1998)]{abrams_1998_nonlinear}
Abrams, D.S.; Lloyd, S.
\newblock Nonlinear {{Quantum Mechanics Implies Polynomial-Time Solution}} for
  {{NP}}-{{Complete}} and \# {{P Problems}}.
\newblock {\em Phys. Rev. Lett.} {\bf 1998}, {\em 81},~3992--3995.
\newblock {\url{https://doi.org/10.1103/PhysRevLett.81.3992}}.

\bibitem[Ferrero et~al.(2004)Ferrero, Salgado, and
  {S{\'a}nchez-G{\'o}mez}]{ferrero_2004_nonlinear}
Ferrero, M.; Salgado, D.; {S{\'a}nchez-G{\'o}mez}, J.L.
\newblock Nonlinear Quantum Evolution Does Not Imply Supraluminal
  Communication.
\newblock {\em Phys. Rev. A} {\bf 2004}, {\em 70},~014101.
\newblock {\url{https://doi.org/10.1103/PhysRevA.70.014101}}.

\bibitem[Rembieli{\'n}ski and Caban(2020)]{rembielinski_2020_nonlinear}
Rembieli{\'n}ski, J.; Caban, P.
\newblock Nonlinear Evolution and Signaling.
\newblock {\em Phys. Rev. Res.} {\bf 2020}, {\em 2},~012027.
\newblock {\url{https://doi.org/10.1103/PhysRevResearch.2.012027}}.

\bibitem[Rembieli{\'n}ski and Caban(2021)]{rembielinski_2021_nonlinear}
Rembieli{\'n}ski, J.; Caban, P. Nonlinear Extension of the Quantum Dynamical Semigroup. {\em Quantum} {\bf 2021}, {\em 5}, 420. 
{\url{https://doi.org/10.22331/q-2021-03-23-420}}.

\bibitem[Kaplan and Rajendran(2022)]{kaplan_2022_causal}
Kaplan, D.E.; Rajendran, S.
\newblock Causal Framework for Nonlinear Quantum Mechanics.
\newblock {\em Phys. Rev. D} {\bf 2022}, {\em 105},~055002.
\newblock {\url{https://doi.org/10.1103/PhysRevD.105.055002}}.

\bibitem[Schr{\"o}dinger(1936)]{schrodinger_1936_probability}
Schr{\"o}dinger, E.
\newblock Probability Relations between Separated Systems.
\newblock {\em Math. Proc. Camb. Philos. Soc.}
  {\bf 1936}, {\em 32},~446--452.
\newblock {\url{https://doi.org/10.1017/S0305004100019137}}.

\bibitem[Park(1968)]{park_1968_nature}
Park, J.L.
\newblock Nature of {{Quantum States}}.
\newblock {\em Am. J. Phys.} {\bf 1968}, {\em 36},~211--226.
\newblock {\url{https://doi.org/10.1119/1.1974484}}.

\bibitem[Park(1988)]{park_1988_thermodynamic}
Park, J.L.
\newblock Thermodynamic Aspects of {{Schr{\"o}dinger}}'s Probability Relations.
\newblock {\em Found. Phys.} {\bf 1988}, {\em 18},~225--244.
\newblock {\url{https://doi.org/10.1007/BF01882932}}.

\bibitem[Beretta(2006)]{beretta_2006_hatsopoulos}
Beretta, G.P.
\newblock The Hatsopoulos--Gyftopoulos Resolution of the Schr{\"o}dinger--Park
  Paradox about the Concept of ``State'' in Quantum Statistical Mechanics.
\newblock {\em Mod. Phys. Lett. A} {\bf 2006}, {\em 21},~2799--2811.
\newblock {\url{https://doi.org/10.1142/S0217732306021840}}.

\bibitem[Beretta(1986)]{beretta_1986_intrinsic}
Beretta, G.P.
\newblock Intrinsic {{Entropy}} and {{Intrinsic Irreversibility}} for a
  {{Single Isolated Constituent}} of {{Matter}}: {{Broader Kinematics}} and
  {{Generalized Nonlinear Dynamics}}. In {\em Frontiers of {{Nonequilibrium
  Statistical Physics}}}; Moore, G.T., Scully, M.O., Eds.; Springer: Boston,
  MA, USA, 1986; pp. 205--212.
\newblock {\url{https://doi.org/10.1007/978-1-4613-2181-1_15}}.

\bibitem[Beretta(1985)]{beretta_1985_entropy}
Beretta, G.P.
\newblock Entropy and Irreversibility for a Single Isolated Two Level System:
  {{New}} Individual Quantum States and New Nonlinear Equation of Motion.
\newblock {\em Int. J. Theor. Phys.} {\bf 1985}, {\em
  24},~119--134.
\newblock {\url{https://doi.org/10.1007/BF00672647}}.

\bibitem[Park and Simmons(1983)]{park_1983_knotsa}
Park, J.L.; Simmons, R.F., The {{Knots}} of {{Quantum Thermodynamics}}.
\newblock In {\em Old and New Questions in Physics, Cosmology, Philosophy, and
  Theoretical Biology: Essays in Honor of Wolfgang Yourgrau}; Springer:
  Boston, MA, USA, 1983; pp. 289--308.
\newblock {\url{https://doi.org/10.1007/978-1-4684-8830-2_20}}.

\bibitem[Hatsopoulos and Beretta(2008)]{hatsopoulos_2008_where}
Hatsopoulos, G.N.; Beretta, G.P.
\newblock Where Is the Entropy Challenge?
\newblock {\em AIP Conf. Proc.} {\bf 2008}, {\em 1033},~34--54.
\newblock {\url{https://doi.org/10.1063/1.2979057}}.

\bibitem[Allahverdyan et~al.(2004)Allahverdyan, Balian, and
  Nieuwenhuizen]{Allahverdyan_2004}
Allahverdyan, A.E.; Balian, R.; Nieuwenhuizen, T.M.
\newblock Maximal work extraction from finite quantum systems.
\newblock {\em Europhys. Lett.} {\bf 2004}, {\em 67},~565.
\newblock {\url{https://doi.org/10.1209/epl/i2004-10101-2}}.

\bibitem[Beretta(1981)]{beretta_1981_thesis}
Beretta, G.P.
\newblock On the general equation of motion of quantum thermodynamics and the distinction between quantal and nonquantal uncertainties (PhD thesis, MIT, 1981). \emph{arXiv} \textbf{2005}, arXiv:quant-ph/0509116. 
\url{https://doi.org/10.48550/arXiv.quant-ph/0509116}.

\bibitem[{Monta{\~n}ez-Barrera} et~al.(2022){Monta{\~n}ez-Barrera}, {von
  Spakovsky}, Damian~Ascencio, and
  {Cano-Andrade}]{montanez-barrera_2022_Decoherencepredictions}
{Monta{\~n}ez-Barrera}, J.A.; {von Spakovsky}, M.R.; Damian~Ascencio, C.E.;
  {Cano-Andrade}, S.
\newblock Decoherence Predictions in a Superconducting Quantum Processor Using
  the Steepest-Entropy-Ascent Quantum Thermodynamics Framework.
\newblock {\em Phys. Rev. A} {\bf 2022}, {\em 106},~032426.
\newblock {\url{https://doi.org/10.1103/PhysRevA.106.032426}}.

\bibitem[Li and von Spakovsky(2016)]{vonspakovsky_2016}
Li, G.; von Spakovsky, M.R.
\newblock Generalized thermodynamic relations for a system experiencing heat
  and mass diffusion in the far-from-equilibrium realm based on steepest
  entropy ascent.
\newblock {\em Phys. Rev. E} {\bf 2016}, {\em 94},~032117.
\newblock {\url{https://doi.org/10.1103/PhysRevE.94.032117}}.

\bibitem[Li et~al.(2018)Li, von Spakovsky, and Hin]{vonspakovsky_2018}
Li, G.; von Spakovsky, M.R.; Hin, C.
\newblock Steepest entropy ascent quantum thermodynamic model of electron and
  phonon transport.
\newblock {\em Phys. Rev. B} {\bf 2018}, {\em 97},~024308.
\newblock {\url{https://doi.org/10.1103/PhysRevB.97.024308}}.

\bibitem[Yamada et~al.(2019)Yamada, von Spakovsky, and
  Reynolds]{vonspakovsky_2019}
Yamada, R.; von Spakovsky, M.R.; Reynolds, W.T.
\newblock Predicting continuous and discontinuous phase decompositions using
  steepest-entropy-ascent quantum thermodynamics.
\newblock {\em Phys. Rev. E} {\bf 2019}, {\em 99},~052121.
\newblock {\url{https://doi.org/10.1103/PhysRevE.99.052121}}.

\bibitem[Monta\~nez Barrera et~al.(2020)Monta\~nez Barrera, Damian-Ascencio,
  von Spakovsky, and Cano-Andrade]{vonspakovsky_2020}
Monta\~nez Barrera, J.A.; Damian-Ascencio, C.E.; von Spakovsky, M.R.;
  Cano-Andrade, S.
\newblock Loss-of-entanglement prediction of a controlled-phase gate in the
  framework of steepest-entropy-ascent quantum thermodynamics.
\newblock {\em Phys. Rev. A} {\bf 2020}, {\em 101},~052336.
\newblock {\url{https://doi.org/10.1103/PhysRevA.101.052336}}.

\bibitem[{von Neumann}(2018)]{vonneumann_2018_mathematical}
{von Neumann}, J.
\newblock {\em Mathematical {{Foundations}} of {{Quantum Mechanics}}}; 
Beyer, R.T., Translator; Princeton University Press: {Princeton, NJ, USA,} 
 2018.
\newblock {\url{https://doi.org/10.2307/j.ctt1wq8zhp}}.

\bibitem[Beretta(2020)]{beretta_2020_fourth}
Beretta, G.P.
\newblock The Fourth Law of Thermodynamics: Steepest Entropy Ascent.
\newblock {\em Philos. Trans. R. Soc. A} {\bf 2020},
  {\em 378},~20190168.
\newblock {\url{https://doi.org/10.1098/rsta.2019.0168}}.

\bibitem[Beretta(2014)]{beretta_2014_steepesta}
Beretta, G.P.
\newblock Steepest Entropy Ascent Model for Far-Nonequilibrium Thermodynamics:
  {{Unified}} Implementation of the Maximum Entropy Production Principle.
\newblock {\em Phys. Rev. E} {\bf 2014}, {\em 90},~042113.
\newblock {\url{https://doi.org/10.1103/PhysRevE.90.042113}}.

\bibitem[Beretta(2019)]{beretta_2019_time}
Beretta, G.P.
\newblock Time--{{Energy}} and {{Time}}--{{Entropy Uncertainty Relations}} in
  {{Nonequilibrium Quantum Thermodynamics}} under {{Steepest-Entropy-Ascent
  Nonlinear Master Equations}}.
\newblock {\em Entropy} {\bf 2019}, {\em 21},~679.
\newblock {\url{https://doi.org/10.3390/e21070679}}.

\bibitem[Murthy et~al.(2023)Murthy, Babakhani, Iniguez, Srednicki, and
  Yunger~Halpern]{YungerHalpern_PRL_2023}
Murthy, C.; Babakhani, A.; Iniguez, F.; Srednicki, M.; Yunger~Halpern, N.
\newblock Non-Abelian Eigenstate Thermalization Hypothesis.
\newblock {\em Phys. Rev. Lett.} {\bf 2023}, {\em 130},~140402.
\newblock {\url{https://doi.org/10.1103/PhysRevLett.130.140402}}.

\bibitem[Simmons and Park(1981)]{simmons_1981_completely}
Simmons, R.F.; Park, J.L.
\newblock On Completely Positive Maps in Generalized Quantum Dynamics.
\newblock {\em Found. Phys.} {\bf 1981}, {\em 11},~47--55.
\newblock {\url{https://doi.org/10.1007/BF00715195}}.

\bibitem[Raggio and Primas(1982)]{raggio_1982_remarks}
Raggio, G.A.; Primas, H.
\newblock Remarks on ``{{On Completely Positive Maps}} in {{Generalized Quantum
  Dynamics}}''.
\newblock {\em Found. Phys.} {\bf 1982}, {\em 12},~433--435.
\newblock {\url{https://doi.org/10.1007/BF00726787}}.

\bibitem[Simmons and Park(1982)]{simmons_1982_another}
Simmons, R.F.; Park, J.L.
\newblock Another Look at Complete Positivity in Generalized Quantum Dynamics:
  {{Reply}} to {{Raggio}} and {{Primas}}.
\newblock {\em Found. Phys.} {\bf 1982}, {\em 12},~437--439.
\newblock {\url{https://doi.org/10.1007/BF00726788}}.

\bibitem[Pechukas(1994)]{pechukas_1994_reduced}
Pechukas, P.
\newblock Reduced {{Dynamics Need Not Be Completely Positive}}.
\newblock {\em Phys. Rev. Lett.} {\bf 1994}, {\em 73},~1060--1062.
\newblock {\url{https://doi.org/10.1103/PhysRevLett.73.1060}}.

\bibitem[Alicki(1995)]{alicki_1995_comment}
Alicki, R.
\newblock Comment on ``{{Reduced Dynamics Need Not Be Completely Positive}}''.
\newblock {\em Phys. Rev. Lett.} {\bf 1995}, {\em 75},~3020--3020.
\newblock {\url{https://doi.org/10.1103/PhysRevLett.75.3020}}.

\bibitem[Pechukas(1995)]{pechukas_1995_pechukas}
Pechukas, P.
\newblock Pechukas {{Replies}}.
\newblock {\em Phys. Rev. Lett.} {\bf 1995}, {\em 75},~3021--3021.
\newblock {\url{https://doi.org/10.1103/PhysRevLett.75.3021}}.

\bibitem[Rivas et~al.(2010)Rivas, Huelga, and Plenio]{rivas_2010_entanglement}
Rivas, {\'A}.; Huelga, S.F.; Plenio, M.B.
\newblock Entanglement and {{Non-Markovianity}} of {{Quantum Evolutions}}.
\newblock {\em Phys. Rev. Lett.} {\bf 2010}, {\em 105},~050403.
\newblock {\url{https://doi.org/10.1103/PhysRevLett.105.050403}}.

\bibitem[Jagadish et~al.(2019)Jagadish, Srikanth, and
  Petruccione]{jagadish_2019_measurea}
Jagadish, V.; Srikanth, R.; Petruccione, F.
\newblock Measure of Positive and Not Completely Positive Single-Qubit
  {{Pauli}} Maps.
\newblock {\em Phys. Rev. A} {\bf 2019}, {\em 99},~022321.
\newblock {\url{https://doi.org/10.1103/PhysRevA.99.022321}}.

\bibitem[Horodecki et~al.(1995)Horodecki, Horodecki, and
  Horodecki]{horodecki_1995_Bell}
Horodecki, R.; Horodecki, P.; Horodecki, M.
\newblock Violating Bell inequality by mixed spin-12 states: necessary and
  sufficient condition.
\newblock {\em Phys. Lett. A} {\bf 1995}, {\em 200},~340--344.
\newblock {\url{https://doi.org/10.1016/0375-9601(95)00214-N}}.

\bibitem[Br{\"u}ning et~al.(2012)Br{\"u}ning, M{\"a}kel{\"a}, Messina, and
  Petruccione]{bruning_2012_parametrizations}
Br{\"u}ning, E.; M{\"a}kel{\"a}, H.; Messina, A.; Petruccione, F.
\newblock Parametrizations of Density Matrices.
\newblock {\em J. Mod. Opt.} {\bf 2012}, {\em 59},~{1--20.} 
\newblock {\url{https://doi.org/10.1080/09500340.2011.632097}}.

\bibitem[Byrd and Khaneja(2003)]{byrd_2003_characterization}
Byrd, M.S.; Khaneja, N.
\newblock Characterization of the Positivity of the Density Matrix in Terms of
  the Coherence Vector Representation.
\newblock {\em Phys. Rev. A} {\bf 2003}, {\em 68},~062322.
\newblock {\url{https://doi.org/10.1103/PhysRevA.68.062322}}.

\bibitem[Horodecki and Horodecki(1996)]{horodecki_1996_informationtheoretic}
Horodecki, R.; Horodecki, M.
\newblock Information-Theoretic Aspects of Inseparability of Mixed States.
\newblock {\em Phys. Rev. A} {\bf 1996}, {\em 54},~1838--1843.
\newblock {\url{https://doi.org/10.1103/PhysRevA.54.1838}}.

\bibitem[Lang and Caves(2010)]{lang_2010_quantum}
Lang, M.D.; Caves, C.M.
\newblock Quantum {{Discord}} and the {{Geometry}} of {{Bell-Diagonal States}}.
\newblock {\em Phys. Rev. Lett.} {\bf 2010}, {\em 105},~150501.
\newblock {\url{https://doi.org/10.1103/PhysRevLett.105.150501}}.

\bibitem[Werner(1989)]{werner_1989_quantum}
Werner, R.F.
\newblock Quantum States with {{Einstein-Podolsky-Rosen}} Correlations
  Admitting a Hidden-Variable Model.
\newblock {\em Phys. Rev. A} {\bf 1989}, {\em 40},~4277--4281.
\newblock {\url{https://doi.org/10.1103/PhysRevA.40.4277}}.

\bibitem[Popescu and Rohrlich(1998)]{popescu_1998_causalitya}
Popescu, S.; Rohrlich, D.
\newblock Causality and {{Nonlocality}} as {{Axioms}} for {{Quantum
  Mechanics}}.
\newblock In \emph{Causality and {Locality} in {Modern
  Physics}}; Hunter, G., Jeffers, S., Vigier, J.P., Eds.; Springer: Dordrecht, The Netherlands,  1998; pp. 383--389.
\newblock {\url{https://doi.org/10.1007/978-94-017-0990-3_45}}.

\bibitem[{Wolfram Research Inc.}(2022)]{wolframresearchinc._2022_mathematica}
{Wolfram Research Inc.} 
\newblock Mathematica {{Online}}.  2022.
\newblock Available online: {\url{https://www.wolfram.com/mathematica} (accessed on 20 July 2023)}. 


\end{thebibliography}

\PublishersNote{}
\end{adjustwidth}
\end{document}